\documentclass[twocolumn, a4paper]{aa}
\usepackage[switch]{lineno}
\usepackage{graphicx}
\usepackage{txfonts}
\usepackage{color}
\usepackage{soul}
\usepackage{natbib}
\usepackage{hyperref}
\hypersetup{
  colorlinks   = true, 
  urlcolor     = blue, 
  linkcolor    = blue, 
  citecolor    = blue  
}
\usepackage{placeins}

\usepackage{amsmath}
\usepackage{float}
\usepackage{multirow}
\usepackage{xcolor}
\usepackage{ulem}

\definecolor{blue}{RGB}{0,0,255}
\definecolor{red}{RGB}{255,0,0}
\definecolor{green}{RGB}{0,255,0}
\definecolor{limegreen}{RGB}{50,205,50}

\begin{document}

\title{Chromospheric and photospheric properties of sunspots as inferred from Stokes inversions under magneto-hydrostatic and non-local-thermodynamic equilibrium}
\author{A.~Vicente Ar\'evalo\inst{1} \and J.M.~Borrero\inst{1} \and I. Mili\'c\inst{1} \and A. Pastor Yabar\inst{2} \and I. Kontogiannis\inst{3,4,5} \and A. G. M. Pietrow\inst{3}}

\institute{Institut f\"ur Sonnenphysik, Georges-K\"ohler-Allee 401A, 79110, Freiburg, Germany
\and
Institute for Solar Physics, Department of Astronomy, Stockholm University, AlbaNova University 
Centre, 10691 Stockholm, Sweden
\and 
Leibniz-Institut f\"ur Astrophysik Potsdam (AIP), An der Sternwarte 16, 14482 Potsdam, Germany 
\and
ETH Z\"urich, Institute for Particle Physics and Astrophysics, Wolfgang-Pauli-Strasse 27, 8093, Zürich, Switzerland
\and 
Istituto ricerche solari Aldo e Cele Dacc\`o (IRSOL), Faculty of Informatics, Universit\`a della Svizzera italiana, CH-6605 Locarno, Switzerland
}

\date{Recieved / Accepted }

\abstract{Sunspots represent a key feature in the solar atmosphere to explore how magnetic fields interact with plasma flows, exhibiting large variations of the physical parameters over very small spatial scales ($< 100$~km), and sometimes featuring dynamic phenomena such as oscillatory umbral flashes.
To fully understand the thermodynamic, magnetic and kinematic structure of these regions, from the stable photosphere to the shock-dominated chromosphere, Stokes inversion techniques are employed to jointly model these layers.}
{We aim to determine the average thermal, magnetic, and kinematic properties of a sunspot from the photosphere to the chromosphere and to deepen our understanding of the properties of umbral flashes.}
{We analyzed high-resolution spectropolarimetric data acquired with the CRISP instrument at the Swedish Solar Telescope (SST). The dataset includes full Stokes measurements of the Mg \textsc{i} 517.2 nm, Na \textsc{i} 589.5 nm, Fe \textsc{i} 630.2 nm, and Ca \textsc{ii} 854.2 nm spectral lines. We performed inversions using the FIRTEZ code, that includes non-local thermodynamic equilibrium (non-LTE) and three-dimensional magneto-hydrostatic (MHS) equilibrium to constrain the gas pressure and density.}
{We successfully inferred the physical parameters in a three-dimensional $(x,y,z)$ domain and provide their average values as a function of the radial distance from the sunspot's center at different heights. Among other findings, we determine that the photospheric Evershed flow is found to reverse into the inverse Evershed inflow in the upper photosphere. In contrast, the moat flow outside the sunspot persists as an outflow at similar heights, suggesting that it is not a direct continuation of the Evershed flow. Furthermore, analysis of an umbral flash event reveals supersonic upflows (Mach numbers $\|M\|\geq 1.5$) and thermodynamic conditions consistent with shock fronts.}
{The application of 3D MHS equilibrium and non-LTE effects combined with multiple lines sensing different layers of the atmosphere allows for the reliable retrieval of atmospheric parameters, which are typically difficult to simultaneously constrain in the photosphere and chromosphere.
The inferred properties of umbral flash show clear evidence of shock dynamics, coinciding with previous theoretical and observational studies that point to converging supersonic flows that move the optical depth iso-surfaces as the driving mechanism behind umbral flashes.} 

\titlerunning{MHS Non-LTE inference of sunspot properties in photosphere and chromosphere}

\authorrunning{Borrero et al.}
\maketitle

\newcommand{\fis}[1]{\textcolor{red}{#1}}
\newcommand{\ava}[1]{\textcolor{blue}{#1}}
\newcommand{\im}[1]{\textcolor{violet}{#1}}
\newcommand{\apy}[1]{\textcolor{orange}{#1}}
\newcommand{\ioa}[1]{\textcolor{olive}{#1}}
\newcommand{\pie}[1]{\textcolor{cyan}{#1}}

\def\kms{~km s$^{-1}$}
\def\deg{^{\circ}}
\def\df{{\rm d}}
\newcommand{\ve}[1]{{\rm\bf {#1}}}
\newcommand{\diff}{{\rm d}}
\newcommand{\Conv}{\mathop{\scalebox{1.5}{\raisebox{-0.2ex}{$\ast$}}}}%
\def\ex{{\bf e_x}}
\def\ez{{\bf e_z}}
\def\ey{{\bf e_y}}
\def\expr{{\bf e_x^\ensuremath{\prime}}}
\def\ezpr{{\bf e_z^\ensuremath{\prime}}}
\def\eypr{{\bf e_y^\ensuremath{\prime}}}
\def\xp{x^\ensuremath{\prime}}
\def\yp{y^\ensuremath{\prime}}
\def\zp{z^\ensuremath{\prime}}
\def\rp{r^\ensuremath{\prime}}
\def\xas{x^{\ast}\!}
\def\yas{y^{\ast}\!}
\def\zas{z^{\ast}\!}
\def\C{\mathcal{C}}
\def\xpp{x^{\prime\prime}}
\def\ypp{y^{\prime\prime}}
\def\zpp{z^{\prime\prime}}

\section{Introduction}
\label{sec:introduction}

\noindent

Sunspots are the most visible manifestation of the solar magnetic field.
Their existence has been known for thousands of years.
From naked-eye observations in ancient China and routine telescopic observations of sunspots recorded for hundreds of years by Harriot, Fabricius, Scheiner, Galileo and others \citep{vaquero2009}, to modern telescopes capable of recording sunspot' polarized spectra with high spatial, spectral, and temporal resolution.
Thanks to advanced analysis techniques such as spectropolarimetry, our knowledge of sunspots has advanced significantly \citep{solanki2003review,borrero2011review,rempel2021review}.\\

Unfortunately, this knowledge pertains mainly to the deepest layers of the solar atmosphere, known as the photosphere \citep{lites1993,stanchfield1997pen,martinez1997pen,mathew2003pen}.
Higher layers such as the chromosphere have been traditionally more difficult to study due to deviations from hydrostatic and local thermodynamic equilibrium (LTE) that hinder the analysis of the spectra and its polarization signals.
Over the past couple of decades the analysis techniques have matured to a point where non-LTE effects \citep{hector2015nicole,jaime2019stic} and magnetohydrostatic equilibrium \citep{borrero2021firtez} can be routinely accounted for, thereby making it possible to study the chromosphere of sunspots \citep{hector2005currents,jaime2013sunspot}.
Despite these important advances, a comprehensive study of the magnetic, kinematic, and thermal properties of sunspots in the chromosphere is still lacking.
This knowledge can have important applications in a number of fields.
It can be used to constrain the upper boundary conditions of three-dimensional magnetohydrodynamic (MHD) simulations of sunspots \citep{rempel2012mhd,jan2020sunspot}, or as a background model where the propagation of MHD waves from the photosphere into the chrosmosphere and their role in heating the upper atmospheric layers can be studied \citep{lena2008sunspot,tobias2012waves,tobias2014waves}.
In addition, it can be used to improve the determination of center-to-limb curves used in the study of stellar activity cycles and exoplanet detection and characterization \citep{lim2023starspot,sumida2026starspot,nemec2026exo,2018AJ....156..189C, 2025A&A...700A.275D}.\\

In the present paper, we employ spectropolarimetric observations of several photospheric and chromospheric spectral lines in a sunspot to determine its average physical properties (e.g. temperature, gas pressure, Wilson depression, magnetic field, line-of-sight velocity, etc.) as a function of the radial distance from the sunspot's center and at various optical depths, ranging from the photosphere to the mid-chromosphere. This is done via the application of a novel Stokes inversion technique \citep{adur2019firtez} that simultaneously accounts for non-local thermodynamic equilibrium effects \citep{kaithakkal2023,borrero2024models} and for the effect of the Lorentz force (i.e. magnetohydrostatic equilibrium) in the thermodynamic structure of the sunspot. In addition to this, we present the results inferred from this inversion technique of a region of the sunspot umbra characterized by the presence of an umbral flash.

\section{Observations}
\label{sec:obs}


The data analyzed in this work was recorded with the CRisp Imaging SpectroPolarimeter (CRISP) instrument \citep{scharmer2008crisp} attached to the 1-meter Swedish Solar Telescope. CRISP was used to measure the Stokes parameters $\ve{I} = (I,Q,U,V)$ across several photospheric and chromospheric spectral lines: Mg {\sc i} 517.2 nm (also known as Fraunhofer's Mg b$_2$ line), Na {\sc i} 589.5 nm (Fraunhofer's Na D$_1$ line), Fe {\sc i} 630.2 nm, and Ca {\sc ii} 854.2 nm.
The atomic data of the recorded spectral lines is summarized in Table~\ref{table:atomicdata}, while the observed wavelengths are given in Table~\ref{tab:scanpos}.\\

The observation strategy followed the one described in \cite{2020A&A...642A.210M}.
The number of accumulations per wavelength and per modulation state was 20 for 854.2nm and 10 for the rest of the spectral windows, and the exposure time per accumulation is 12 miliseconds.
Part of data processing consisted of image reconstruction using the Multi-Object Multi-Frame Blind Deconvolution technique \cite[MOMFBD][]{lofhadl2002momfbd,michiel2005momfbd} to remove the effects of atmospheric seeing during data acquisition and reach diffraction-limited observations.
Given that the diffraction limit varies with wavelength, the data were first co-aligned and de-scretched using wide-band data for each spectral window and taking as a reference the shortest wavelength (517.2 nm).
The narrow-band data were first corrected for cavity errors (small wavelength shifts) and then downgraded to the largest spatial sampling (589.6 nm), where the pixel scale is 0.0446 $\arcsec$~pix$^{-1}$.
The size of the observed field of view is about 59$\arcsec$ $\times$ 59$\arcsec$ sampled in a total of 1358$\times$1321 pixels.\\

The noise level achieved in the polarization signals is estimated by calculating the standard deviation of the polarization profiles at a fixed wavelength position. The wavelength used is in our case $-0.6$ {\AA} from the center of the Fe {\sc I} line, where no signal should be present in $Q$, $U$ and $V$.
This yields a noise level of $\sigma_{\rm pol} \approx 1.6 \times 10^{-3}$. The data described until here will be referred to as diffraction-limited data.\\

\begin{table*}
\begin{center}
\caption{Atomic parameters of the spectral lines analyzed in this work}
\begin{tabular}{ccccccccc}
\hline
Ion & $\lambda_{0}$ & $\chi_{\rm low}$ & $\log(gf)$ & Elec.conf & $\sigma$ & $\alpha$ & $g_{\rm eff}$ & $G_{\rm lin}$\\
 & [{nm}] & [eV] & & & & & & \\
\hline
Mg \textsc{i}  & 517.26843 & 2.711 & -0.393 & ${^3}P_{1}-{^3}P_{1}$ & 729 & 0.238 & 1.5 & 2.25\\
Na \textsc{i}  & 589.59242 & 0.000 & -0.194 & $^2$S$_{1/2}-{^2}$P$_{1/2}$ & 407 & 0.273 & 1.33 & 2.22\\
Fe \textsc{i}  & 630.24936 & 3.686 & -1.235 & $^5$P$_1-{^5}$D$_0$ & 856 & 0.240 & 2.5 & 10.62\\
Ca \textsc{ii} & 854.20900 & 1.700 & -0.360 & ${^2}D_{5/2}-{^2}P_{3/2}$ & 291 & 0.275 & 1.10 & 1.27\\
\hline
\end{tabular}
\tablefoot{$\lambda_{0}$ represents the central laboratory wavelength of the spectral line. $\sigma$ and $\alpha$ represent the cross-section (in units of Bohr's radius squared $a_0^2$) and velocity parameter of the atom undergoing the transition, respectively, for collisions with neutral atoms under the ABO theory \citep{abo1,abo2,abo3}. The effective Land\'e factor $g_{\rm eff}$ has been calculated under the assumption of LS coupling. The Land\'e factor for the linear polarization $G_{\rm lin}$ has been calculated according to \citet{egidio92book}.\label{table:atomicdata}}
\end{center}
\end{table*}

In addition to the diffraction-limited data, in this work we also analyze spatially averaged data where the diffraction-limited data is binned in a 4$\times$4 pixel region, resulting in a pixel scale of 0.176$\arcsec$~pix$^{-1}$ and a total of 339$\times$330 pixels on the solar surface $(x,y)$. We estimate the noise level in the polarization signals of the binned data to be $\sigma_{\rm pol} \approx 1.2 \times 10^{-3}$. The noise reduction achieved via binning deviates from the expected $\propto \sqrt{1/N}$ scaling. This occurs because the noise in the diffraction-limited data is spatially correlated, a consequence of the Wiener filter applied during the MOMFBD image reconstruction, which amplifies low-frequency noise components while suppressing high-frequency power, thus resulting in a lower than expected noise reduction.\\

\begin{table*}
\begin{center}
\caption{Scanning positions of the observed spectral lines.\label{tab:scanpos}}
\begin{tabular}{ccccl}
\hline
Line & N$_{\lambda,{\rm syn}}$ & $\Delta\lambda_{\rm syn}$ & N$_{\lambda,{\rm obs}}$ & observed wavelengths $10^{-3}$nm \\
\hline
Mg \textsc{i}  & 45 & $3\cdot10^{-3}$nm & 13 & 0, $\pm 3$, $\pm 6$, $\pm 9$, $\pm 12$, $\pm 24$, $\pm 65$ \\
Na \textsc{i}  & 35 & $7\cdot10^{-3}$nm & 13 & 0, $\pm 7$, $\pm 14$, $\pm 24$, $\pm 34$, $\pm 6$, $\pm 120$ \\
Fe \textsc{i}  & 21 & $6\cdot10^{-3}$nm & 11 & 0, $\pm 6$, $\pm 12$, $\pm 18$, $\pm 30$, $\pm 60$ \\
Ca \textsc{ii} & 81 & $6\cdot10^{-3}$nm & 21 & 0, $\pm 3$, $\pm 9$, $\pm 15$, $\pm 21$, $\pm 27$, $\pm 033$, $\pm 39$, $\pm 45$, $\pm 51$, $\pm 240$ \\
\hline
\end{tabular}
\tablefoot{N$_{\lambda,{\rm syn}}$ is the number of synthesized wavelengths for each line, separated by $\Delta\lambda_{\rm syn}$ $10^{-3}$nm. 
N$_{\lambda,{\rm obs}}$ is the number of observed wavelengths, distributed around the center one as the observed wavelengths show.}
\end{center}
\end{table*}

The target recorded with the CRISP instrument corresponds to NOAA AR 13433 and was observed on September 15th, 2023 at 8:38 UT.
At this time, the center of the sunspot was located at approximate coordinates (-458$\arcsec$,355$\arcsec$) corresponding to a heliocentric angle $\Theta \approx 39.4^{\circ}$ (i.e., $\mu\approx0.77$).
Maps of the intensities observed at different wavelengths are shown in the top panels of Figure~\ref{fig:results_tau_1}.
The first one corresponds to the continuum intensity at 630 nm, whereas the second and third maps show the observed intensity at the line core position of the Mg {\sc i} and Ca {\sc ii} lines, respectively.
In this order, they are meant to represent three different layers of the solar atmosphere: the photospheric continuum, the upper photosphere, and the chromosphere sampled by these spectral lines (see also the Appendix~\ref{app:response_functions}).
Figures~\ref{fig:results_tau_1} and ~\ref{fig:results_tau_2} contain three blue concentric circles around the sunspot's center (indicated by the red cross in the umbra). These circles are meant to approximately represent the umbra-penumbra boundary, penumbra-quiet Sun boundary, and superpenumbral boundary, respectively. They are located around $r/R_{\rm s} = 0.3, 1.0$ and $1.4$, respectively, where $R_{\rm s}$ represents the sunspot's radius. In our case $R_{\rm s} = 22\arcsec$.\\

\section{Stokes inversion and results}

\subsection{Inversion procedure}
\label{subsec:inversion}

The Stokes inversion code used in this work is the FIRTEZ code \citep{adur2019firtez,borrero2019firtez}.
This code derives the physical parameters in the three-dimensional domain $ (x, y, z) $, representing the solar atmosphere, from observations of the Stokes measurements on the solar surface $\ve{I}(x,y,\lambda)$. The simultaneous inversion of multiple spectral lines that form under potentially different physical conditions, such as local versus non-local thermodynamic equilibrium (LTE vs NLTE), and at different heights, while enforcing physical constraints like magneto-hydrostatic equilibrium (MHS), is a delicate task that deserves special attention.\\

Although the general procedure has been described in detail in \citet{borrero2024models}, here we provide a short summary and some additional details when the approach deviates from the one described in the aforementioned paper.
In this context, it is important to mention that for this work, the NLTE treatment in FIRTEZ has been improved by separating the source function into its contributions: one arising from the continuum and another contribution arising from the atomic transition (i.e. spectral line), with only the latter being affected by the departure coefficients.
These modifications are detailed in Appendix~\ref{app:source_nlte}.\\

To initialize the inversion, we first obtain a guess of the physical parameters.
To this end, we used the temperature and gas pressure from the VALC model \citep{vernazza1981} interpolated to 128 vertical grid points with a spacing of $\Delta z = 12$km ($\Delta\log\tau_{\rm c} \approx 10^{-1}$ in the deep photosphere). 
The initial estimates for the magnetic field and line-of-sight velocity are done using simple estimations via the weak field approximation and the center of gravity method, respectively, and derived from the Fe {\sc i} line at 630 nm only.
Consequently, the initial estimate for these two parameters has constant values with optical depth.
The VALC model is also used to compute initial departure coefficients for the Mg {\sc i}, Na {\sc i}, and Ca {\sc ii} spectral lines (see Figure~\ref{fig:dep_coeffs}).
More information on the determination of the departure coefficients can be found in Appendix~\ref{app:betas}. The Fe {\sc i} line at 630 nm is always considered to be formed under LTE conditions. The inversion procedure is divided into several inversion cycles. During each inversion cycle, the departure coefficients are kept constant. This approach is similar to the so-called Fixed Departure Coefficient approximation \citep[FDC;][]{hector1998nlte}. All inversion cycles described hereafter were carried out by giving equal weights to the four Stokes parameters $(I,Q,U,V)$.\\

With these ingredients, a first inversion cycle of the spatially binned data (see the description in Sect.~\ref{sec:obs}) is performed employing vertical (i.e., one-dimensional) hydrostatic equilibrium. Four nodes were used for each of the following physical parameters: temperature $T$, line-of-sight velocity $v_{\rm los}$, micro-turbulent velocity $v_{\rm mic}$, and the three components of the magnetic field ($B_x$, $B_y$, $B_z$). This represents a total of 24 free parameters that are used to fit a total of 232 observed wavelengths ($N_{\lambda,\rm obs}=58$ for each of the four Stokes parameters; see Table~\ref{tab:scanpos}).\\

After the first inversion cycle, the resulting components of the magnetic field perpendicular to the line-of-sight ($B_x$, $B_y$) are then corrected from the 180$^{\circ}$ ambiguity using the Non-Potential Field Calculation method \citep{georgoulis2005}. Next, the gas pressure is recalculated under the assumption of three-dimensional (3D) magneto-hydrostatic equilibrium (MHS; i.e., taking into account the Lorentz force) as described in \citet{borrero2021firtez}. Because the temperature and gas pressure have changed through the application of the Stokes inversion and MHS equilibrium, some of the departure coefficients (obtained for a VALC model) are no longer representative of the new physical conditions. Therefore, in principle they should be recalculated for every point on the three-dimensional domain and for each spectral line formed under NLTE conditions. In practice, however, this is done only for the Ca {\sc ii} line, with the departure coefficients for Mg {\sc i} and Na {\sc i} being kept the same as determined initially for the VALC model. This decision is justified in detail in Appendix~\ref{app:betas}. After this, the fit to the observed Stokes profiles is no longer adequate, and therefore a new inversion cycle of the spatially binned data is performed (keeping the gas pressure constant to what the MHS equilibrium establishes) to make sure the new physical parameters adapt to the change in the departure coefficients. The number of nodes is now reduced to: 3 for $T$, $v_{\rm los}$ and $v_{\rm mic}$, and two for $B_x$, $B_y$, and $B_z$ (15 in total).\\

After the second inversion cycle, we obtain: temperature $T(x,y,z)$, line-of-sight velocity $v_{\rm los}(x,y,z)$, magnetic field $\ve{B}(x,y,z)$, gas pressure $P_{\rm g}$, and density $\rho (x,y,z)$.
We emphasize that, because FIRTEZ determines the gas pressure using three dimensional MHS equilibrium instead of vertical hydrostatic equilibrium, the physical parameters are inferred in the $(x,y,z)$ domain in addition to the more commonly used $(x,y,\tau_{\rm c})$, where $\tau_{\rm c}$ refers to the optical depth \citep{borrero2021firtez}. The conversion between $z$ and $\tau_{\rm c}$ is non-linear and therefore surfaces of constant optical depth do not correspond to surfaces of constant $z$, as can be seen in panel f of Fig. \ref{fig:radial_averages} and on the vertical cut of Fig. \ref{fig:temperature_velocity_umbral_flash_cut}. \\

\begin{figure}[htbp]
    \centering
    \includegraphics[width=1\linewidth]{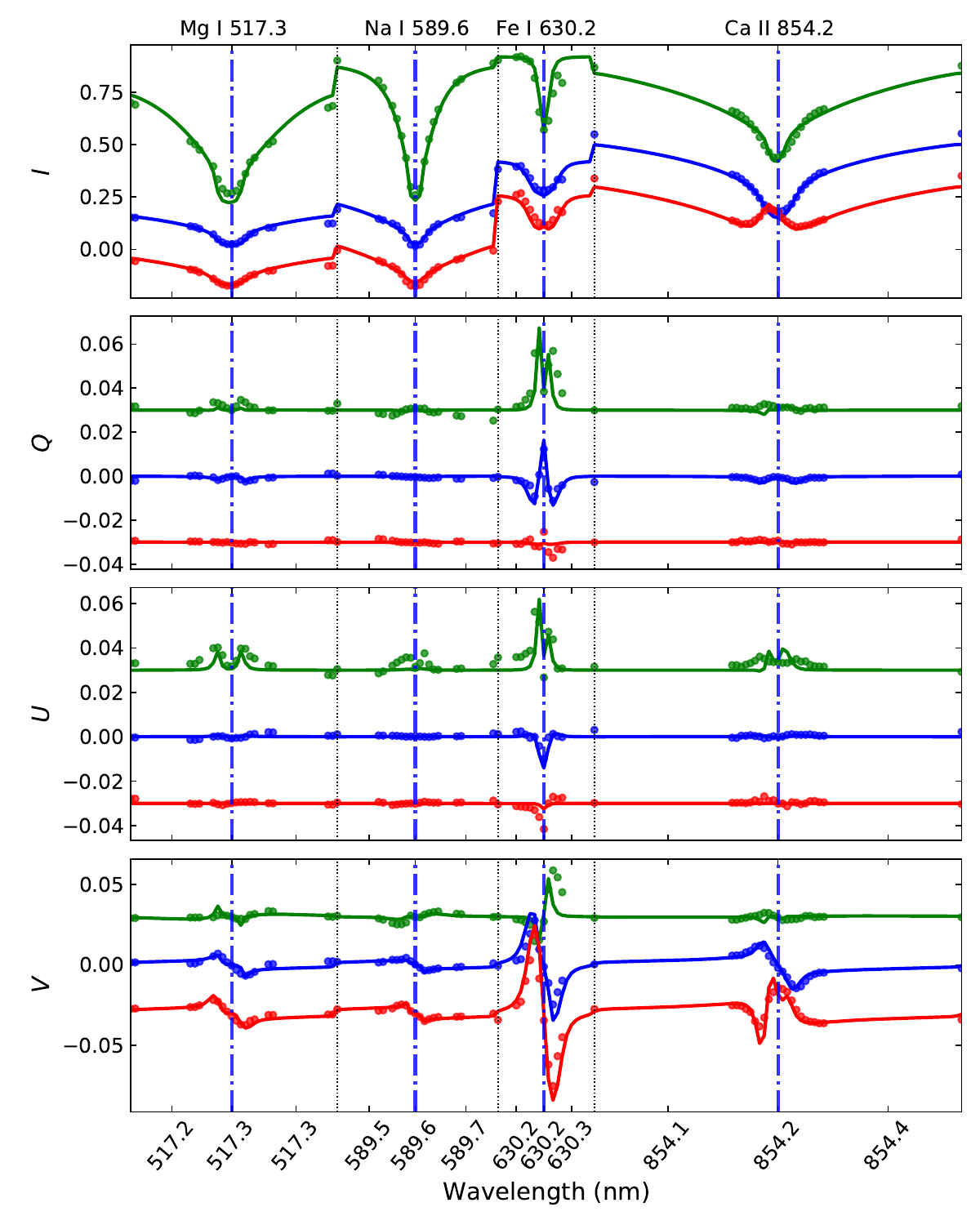}
    \caption{Representative Stokes profiles at three different locations in the diffraction limited observations. The three locations are: penumbra (green), umbra (blue), umbral flash (red). The points display the observed data whereas solid lines correspond to the fitted profiles.
    The profiles are normalized to the quiet sun continiuum and in order to
    better visualize the data, vertical shifts have been applied. These are: $[+0.2, 0, -0.2]$ in Stokes $I$ and [+0.02, 0, -0.02] in Stokes $Q$, $U$ and $V$. In addition, Stokes $I$ signals in the umbra (blue) and umbral flash (red) have been multiplied by a factor 2 before the shift applied to better display the variations.
    }
    \label{fig:sample_fits}
\end{figure}

At this point, we remind that the procedure described above applies to the inversion of the spatially binned data (see Section~\ref{sec:obs}). The inversion of the diffraction-limited data is done by reinterpolating the results of the inversion of the binned data (330$\times$330 pixels) to the original resolution (1321$\times$1358 pixels) and using this as an initial guess in a new inversion cycle that fits the diffraction-limited data. This new inversion is done with the same number of nodes or free parameters as in the second cycle of the inversion of the spatially binned data: 3 for $T$, $v_{\rm los}$ and $v_{\rm mic}$ and two for $B_x$, $B_y$ and $B_z$. The reason for doing this is two fold. First, it helps save computation time, as the initial inversion steps, MHS determination of the gas pressure, and recalculation of departure coefficients are done at a lower resolution (i.e. a lower number of spatial pixels). Second, the lower noise level in the spatially binned data helps obtain smoother maps that are then employed to initialize the inversion of the diffraction limited data. This significantly improves the results for the components of the magnetic field perpendicular to the observer's line-of-sight $(B_x, B_y)$ that are inferred from the, typically low signals, in the linear polarization Stokes $Q$ and $U$. This also improves the subsequent 180$^{\circ}$ disambiguation and therefore also the determination of the Lorentz force and gas pressure via magneto-hydrostatic equilibrium. Examples of representative line profiles of the observations along with its fitted profiles for the difraction limited data are shown in Fig. \ref{fig:sample_fits}.

\subsection{Inversion results}
\label{subsec:results}

As a result of the Stokes inversion described in Sect.~\ref{subsec:inversion} the FIRTEZ code provides the following physical parameters: temperature $T$, magnetic field vector $\ve{B}$, line-of-sight velocity $v_{\rm los}$, micro-turbulent velocity $v_{\rm mic}$, gas pressure $P_{\rm g}$, density $\rho$, and continuum optical depth $\tau_{\rm c}$
with its corresponding uncertainties. The uncertanties were computed using the inverse of the Hessian matrix \citep[see Eq.~4 in][]{borrero2024firtez}.
Although this method can differ from the true error of the complete inversion process (i.e, it neglects degeneracies and local minima), over(under)estimating the true error depending on the relative strength of the neglected effects, it is sufficient to give us a proxy of how narrow is the $\chi^2$ valley in the inversion process.

All physical parameters are provided in a $(x,y,z)$ three dimensional domain that is a combination of the $(x,y)$ domain from the observed Stokes parameters (see Sect.~\ref{sec:obs}) and the $z$ domain from the initial 1D guess model (VALC; see Sect.~\ref{subsec:inversion}).
Consequently, the inversion of the spatially binned data in the three dimensional domain has $n_x = 339$, $n_y=330$, $n_z=128$ points, with the following spacing along each of those dimensions: $\Delta x = \Delta y= 176$~km, $\Delta z=12$~km. On the other hand, the three-dimensional domain in the inversion of the diffraction limited data has  $n_x = 1358$, $n_y=1321$, $n_z=128$ points with spacings of $\Delta x = \Delta y= 44$~km, $\Delta z=12$~km. To illustrate how the vertical height $z$ maps to the more commonly used optical depth $\tau_{\rm c}$ scale, we show some hints at panel f of Fig. \ref{fig:radial_averages} and in the vertical cut of Fig. \ref{fig:temperature_velocity_umbral_flash_cut}.
In the remaining of this section, and in Section~\ref{sec:properties} the results from the inversion of the spatially binned data will be used. Section~\ref{sec:flash} will employ the results from the inversion of the diffraction limited data.

\begin{figure*}[!htp]
\begin{tabular}{ccc}
\includegraphics[width=5.8cm]{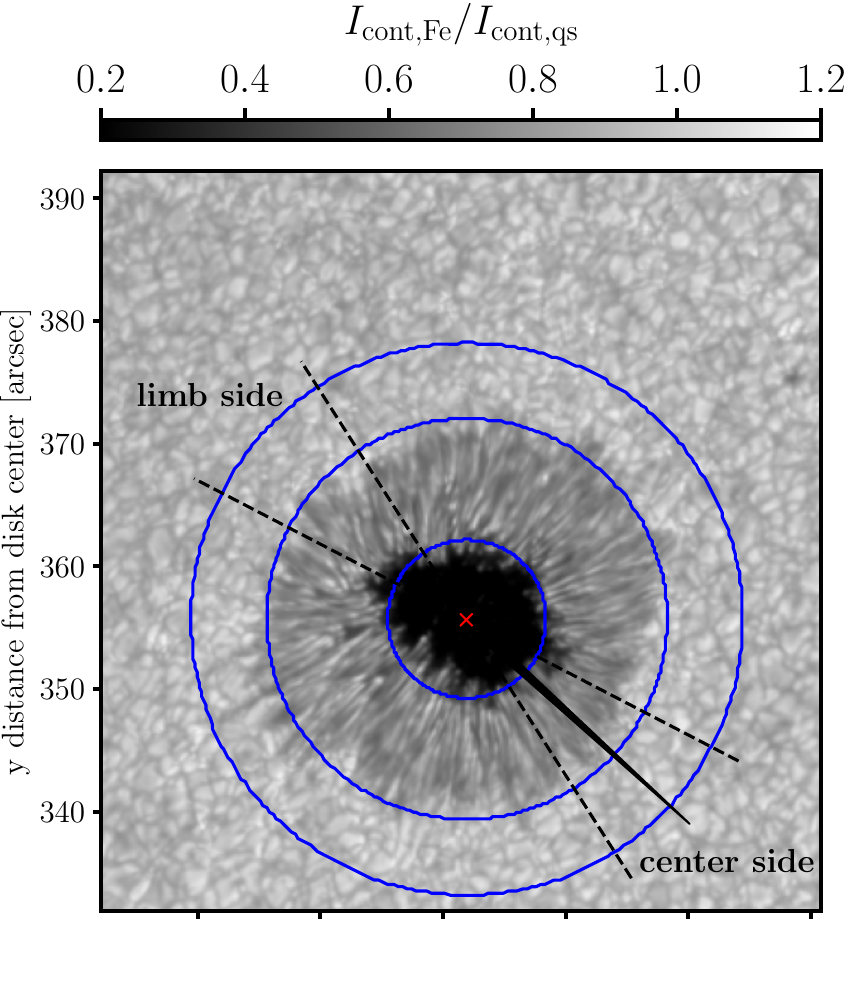} &
\includegraphics[width=5.8cm]{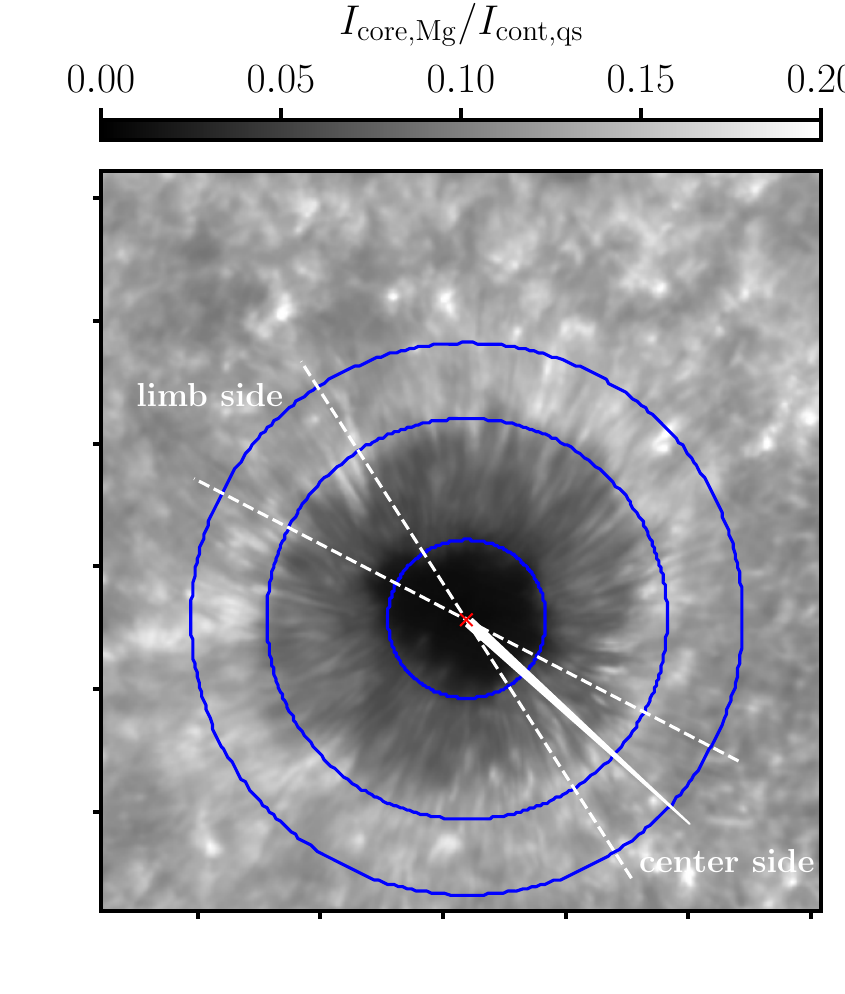} &
\includegraphics[width=5.8cm]{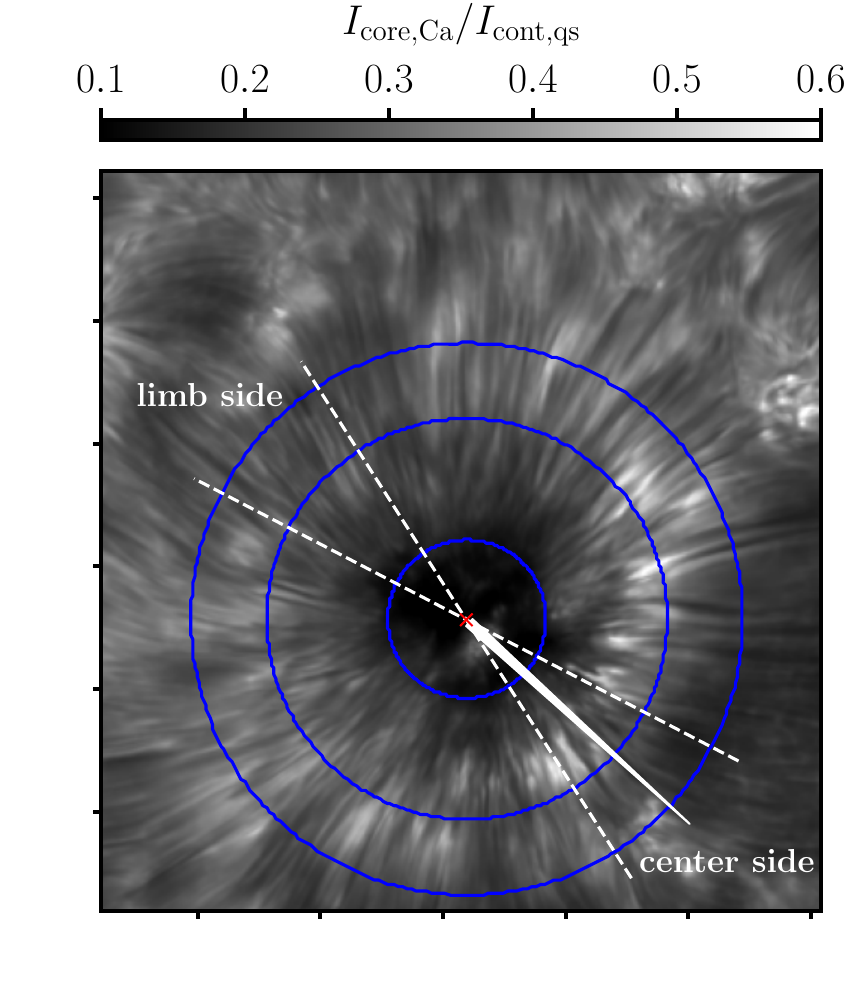}\\
\includegraphics[width=5.8cm]{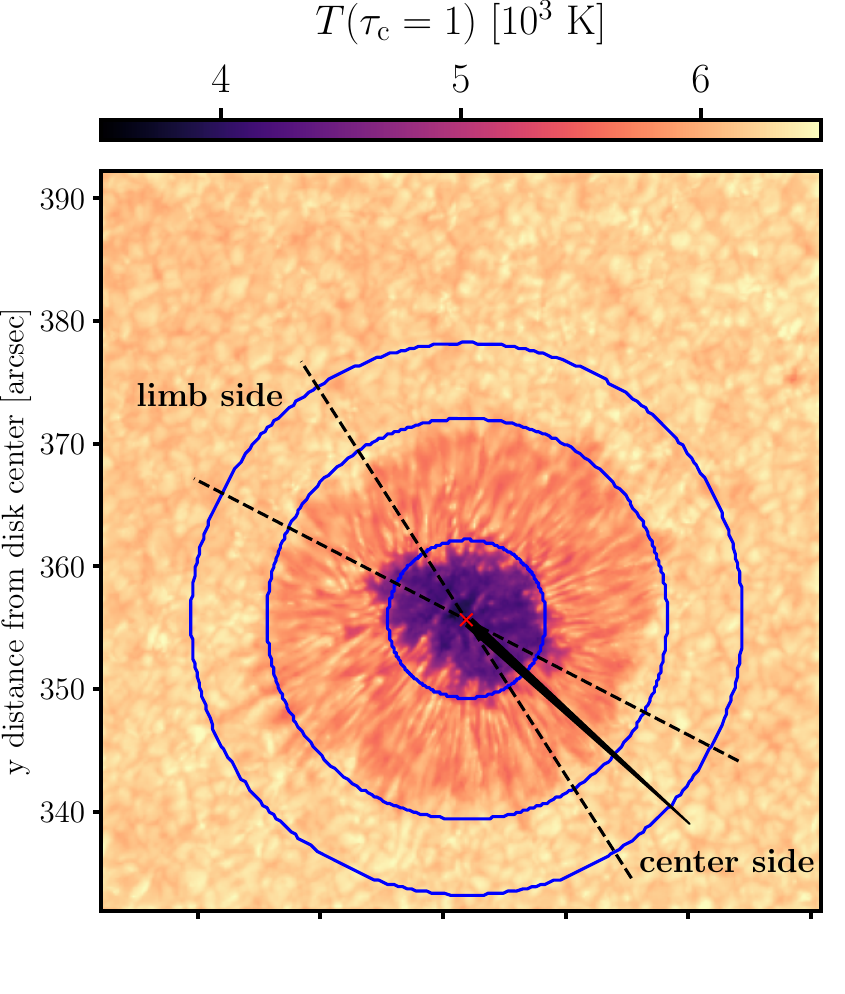} &
\includegraphics[width=5.8cm]{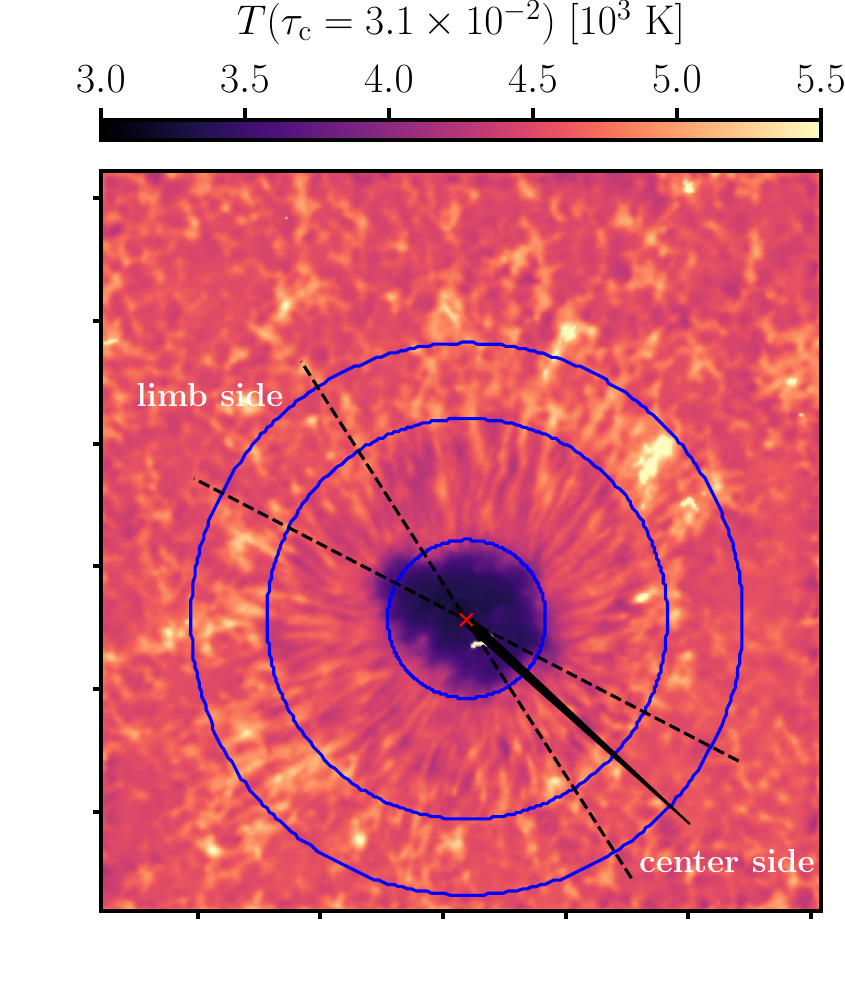} &
\includegraphics[width=5.8cm]{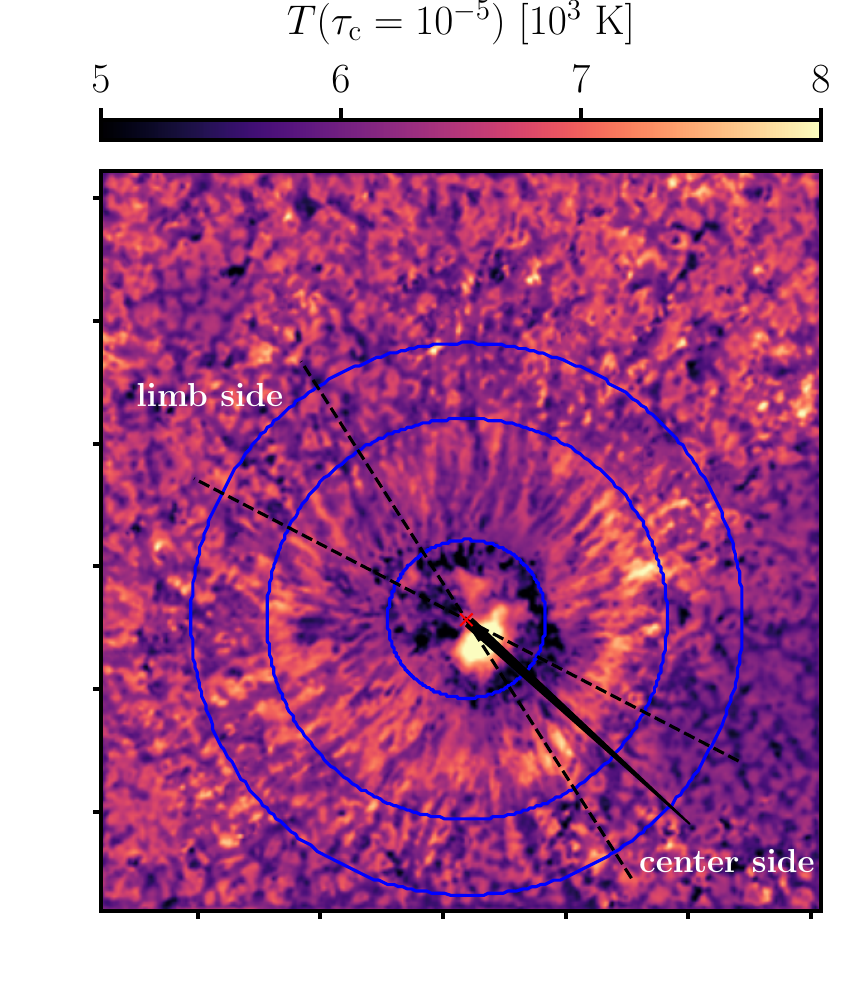}\\
\includegraphics[width=5.8cm]{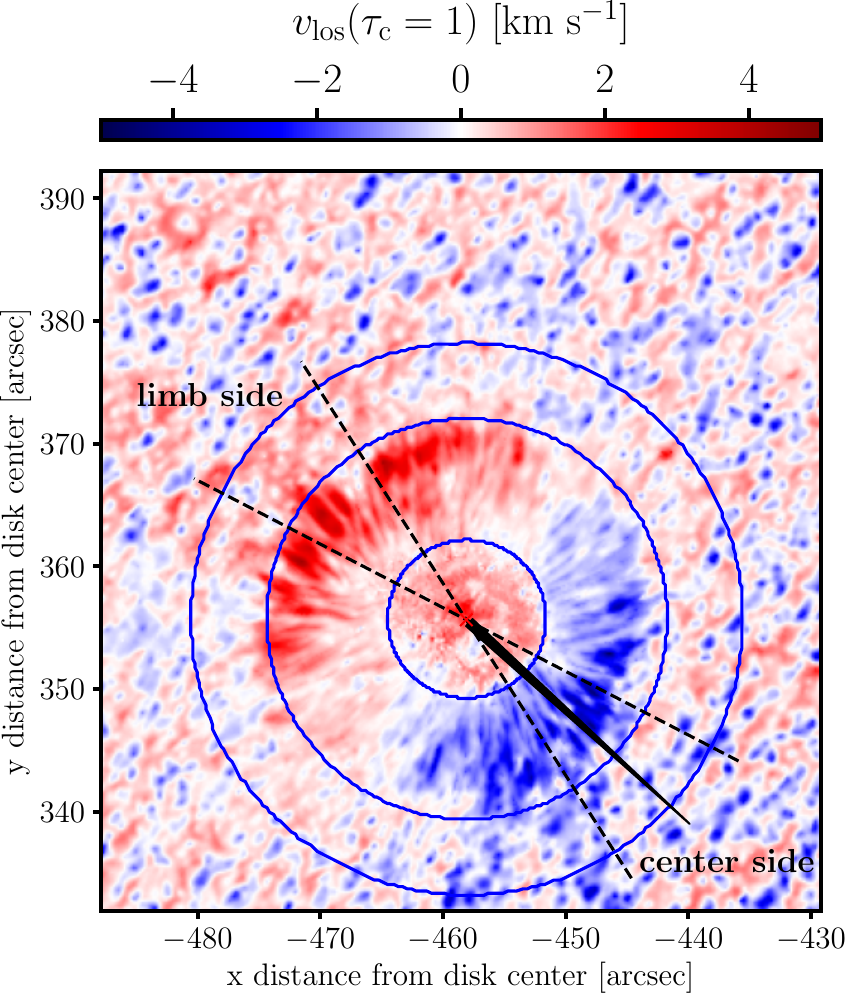} &
\includegraphics[width=5.8cm]{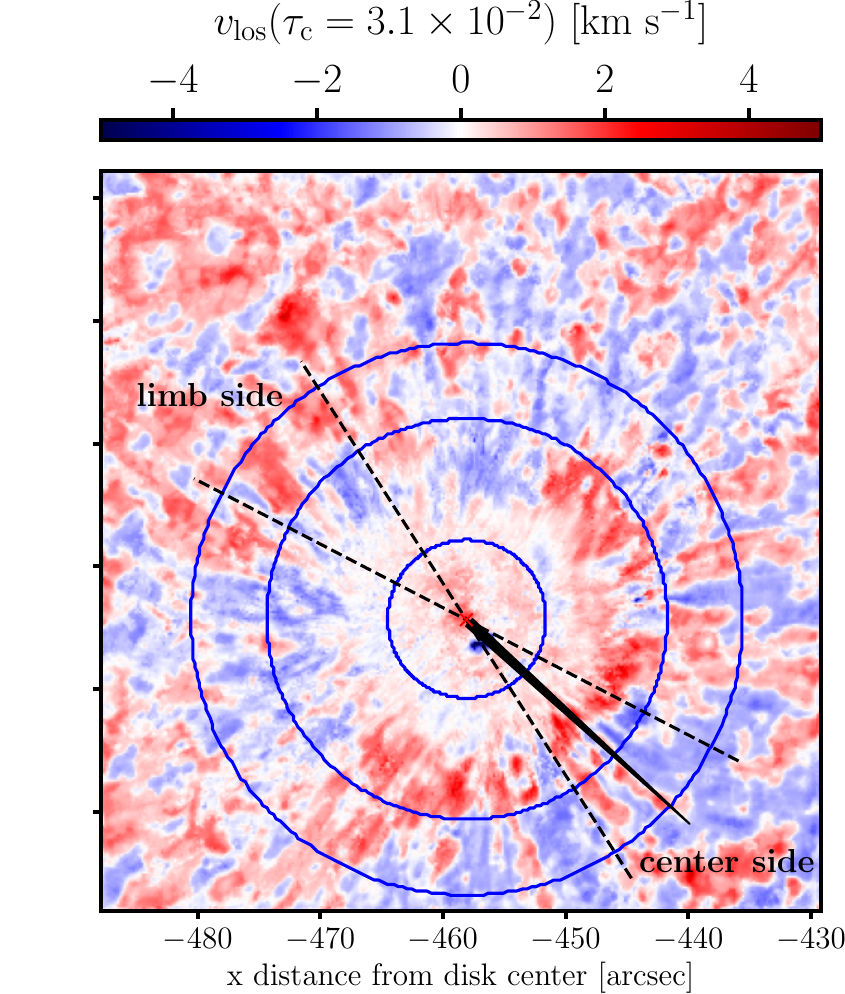} &
\includegraphics[width=5.8cm]{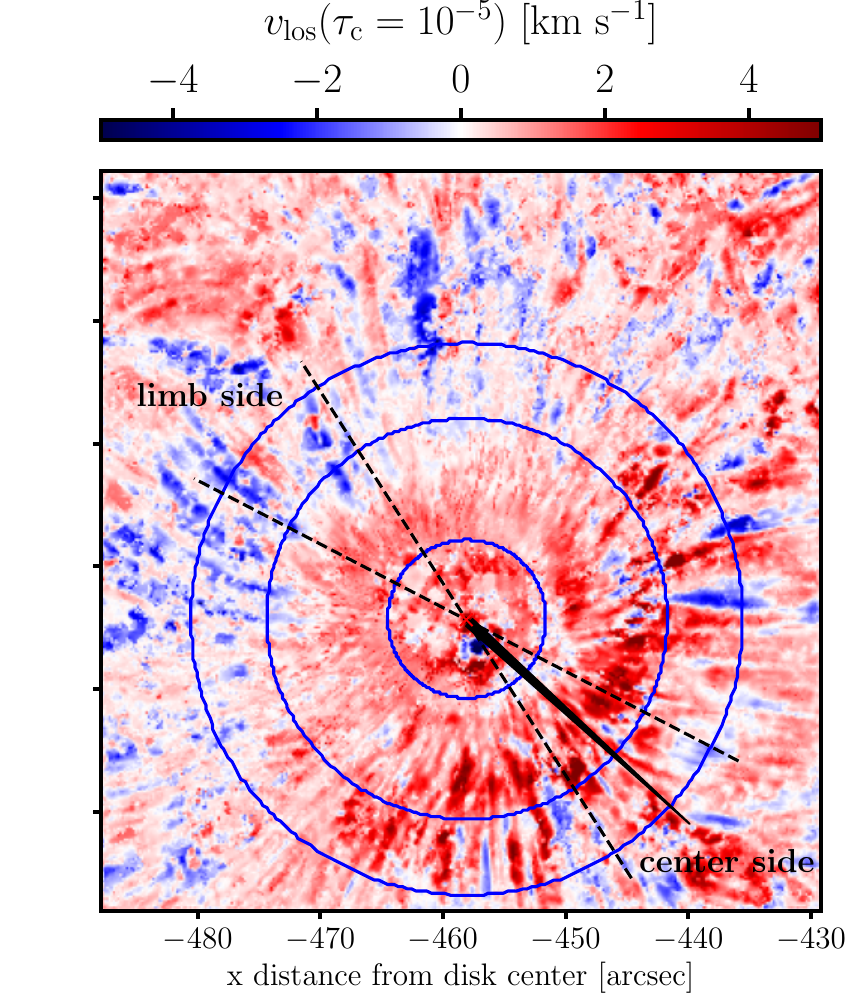}\\
\end{tabular}
\begin{center}
\caption{Observations and inversion results. Top panels display maps of the observed intensity at three different wavelengths (from left to right): continuum of the Fe {\sc i} line at 630 nm, core of the Mg {\sc i} line at 517 nm, and core of the Ca {\sc ii} line at 854 nm. Middle panels: temperature inferred from the inversion at three different optical depth levels (from left to right): $\tau_{\rm c}=1$ (photospheric continuum), $\tau_{\rm c}=10^{-2.5}$ (high photosphere), and $\tau_{\rm c}=10^{-5}$ (chromosphere). Bottom panels: same as middle panels but showing the inferred line-of-sight velocity. Inner and outer blue circles are located at radii $r/R_{\rm s} = 0.4, 1.0, 1.4$, respectively, where $R_{\rm s}$ is defined as the sunspot's radius as measured from the umbral center (red cross). The direction of the solar disk center is indicated by the bold arrow. Black dashed lines depict a cone of $\pm 15^{\circ}$ around the line of symmetry of the spot.\label{fig:results_tau_1}}
\end{center}
\end{figure*}

\subsubsection{Temperature}
\label{subsubsec:temperature}

Figure~\ref{fig:results_tau_1} (middle row panels) show the inferred temperature $T$ as a function of $(x,y)$ for three different surfaces of constant optical depth: $\tau_{\rm c}=1$ (i.e. deep photosphere), $\tau_{\rm c}=10^{-2.5}$ (i.e. high photosphere), and $\tau_{\rm c}=10^{-5}$ (chromosphere). As demonstrated in App.~\ref{app:response_functions}, our observed spectral lines provide us with reliable information in all these regions. As can be seen, the temperature maps in the middle panels of Fig.~\ref{fig:results_tau_1} display a very high correlation with the observed intensity levels at the three different wavelengths presented in the upper panels of the same figure.\\

\begin{figure*}[ht!]
\begin{tabular}{ccc}
\includegraphics[width=5.8cm]{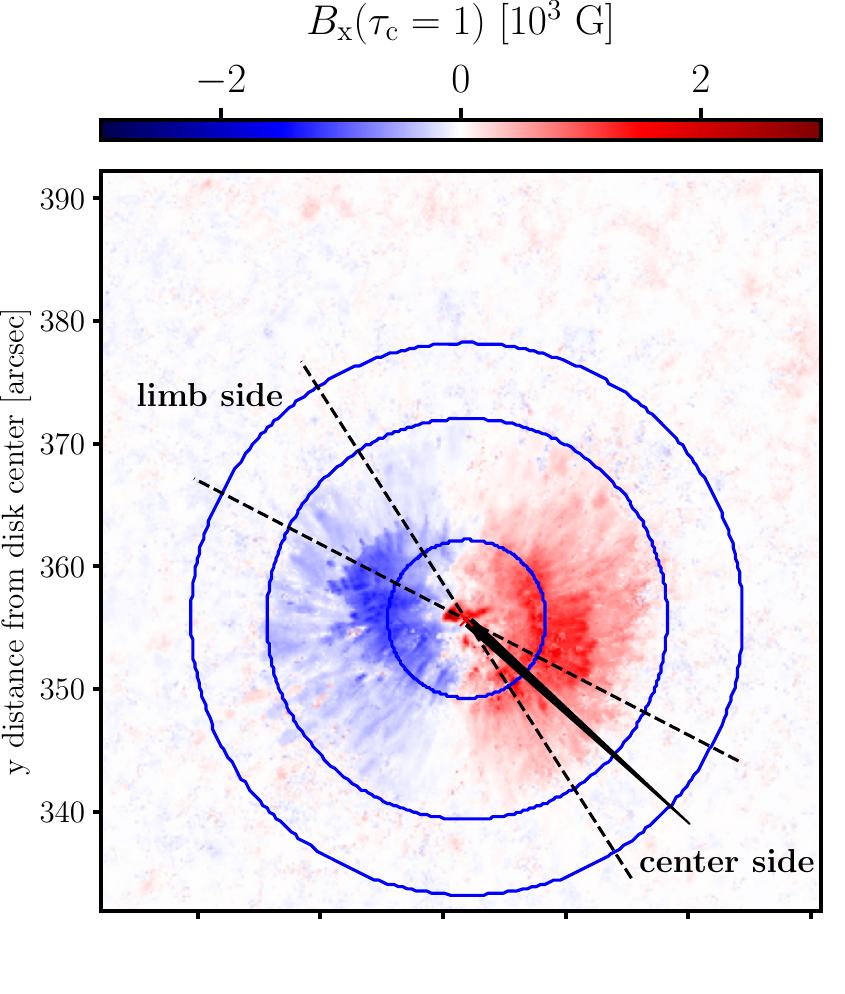} &
\includegraphics[width=5.8cm]{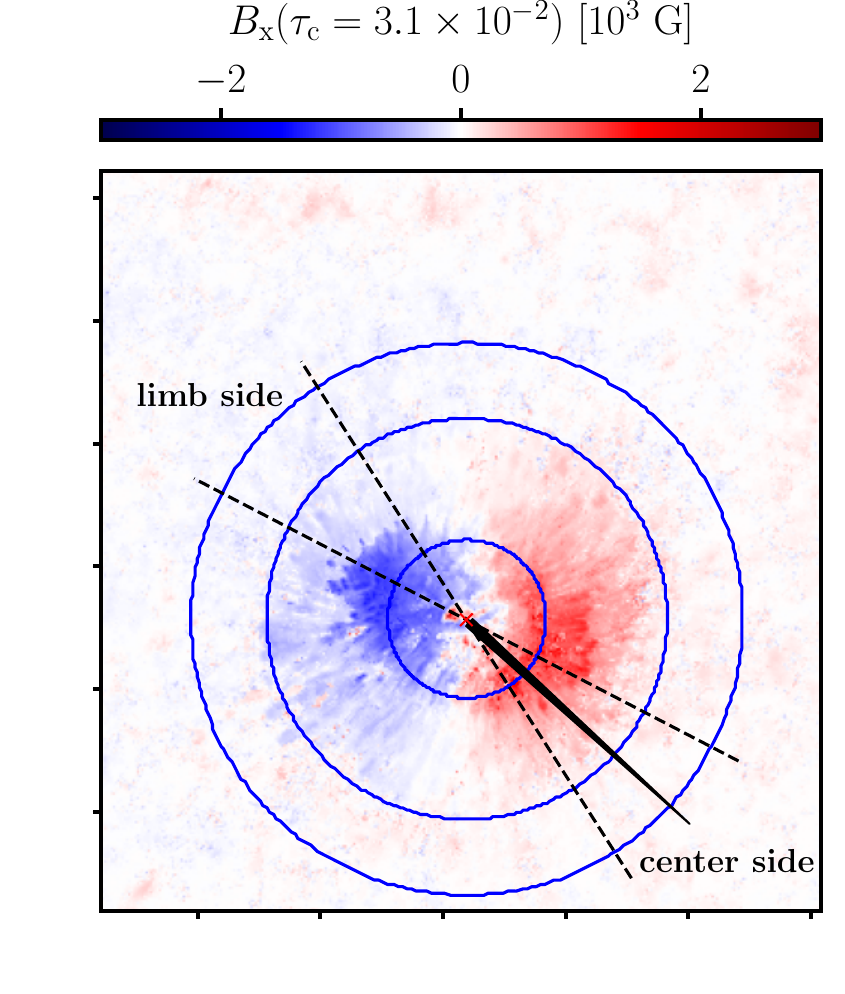} &
\includegraphics[width=5.8cm]{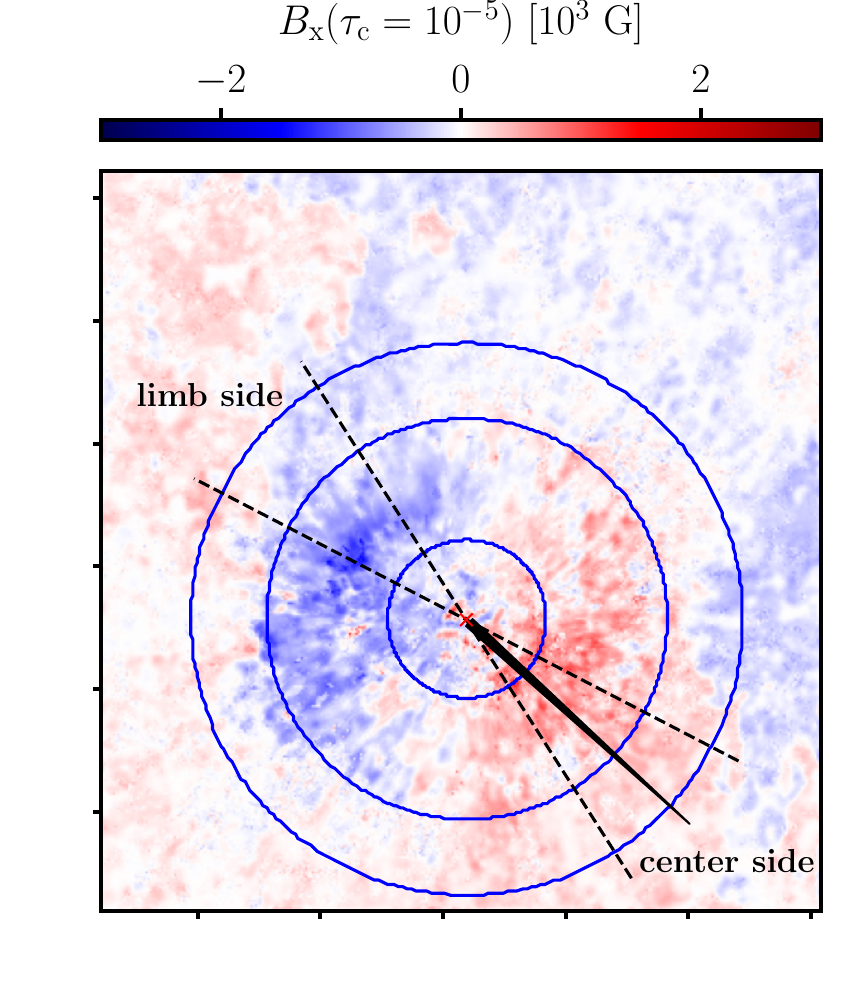}\\
\includegraphics[width=5.8cm]{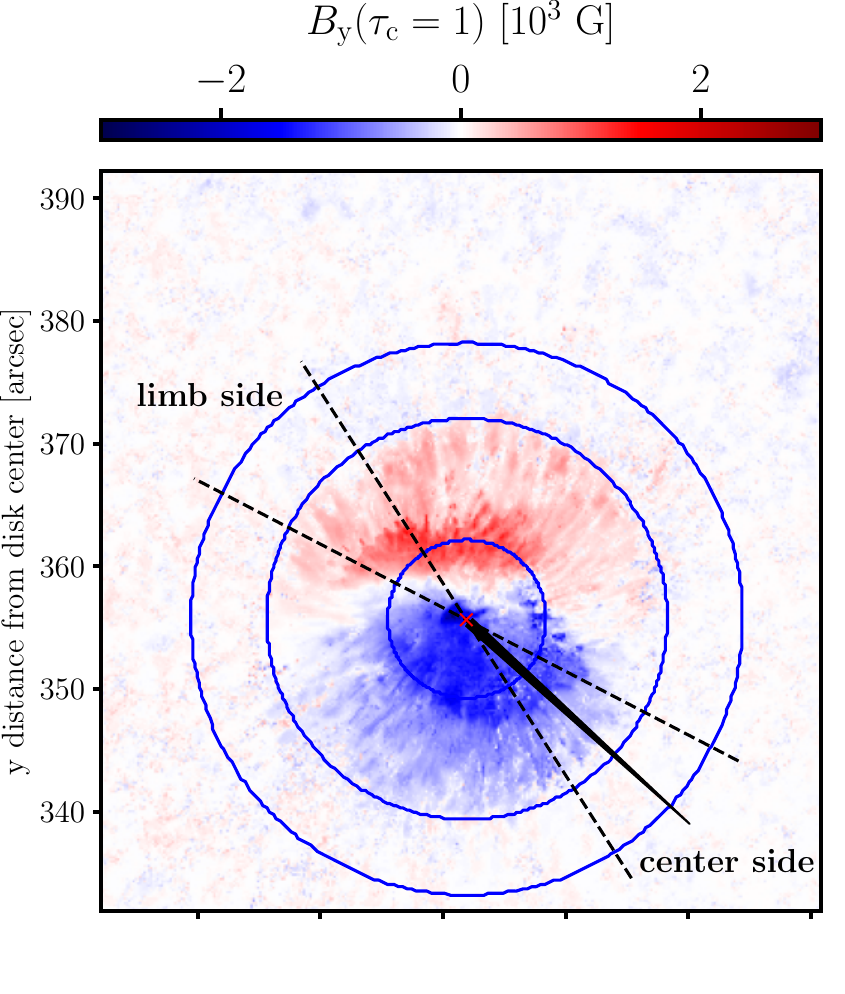} &
\includegraphics[width=5.8cm]{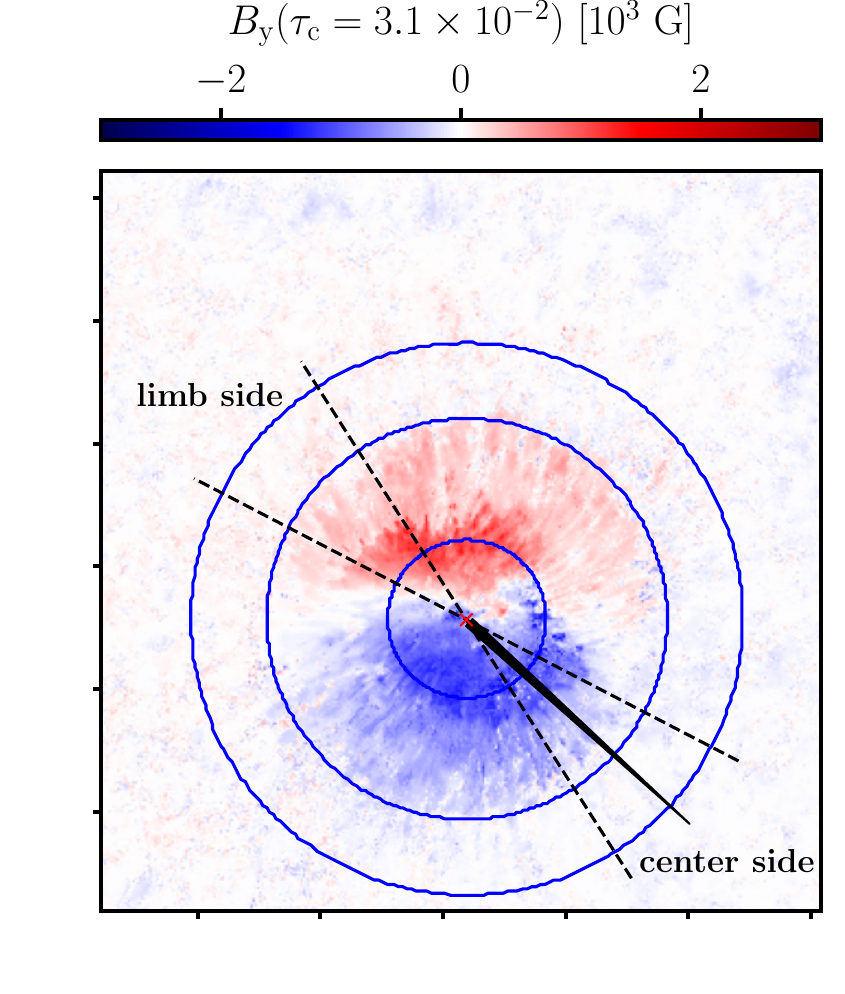} &
\includegraphics[width=5.8cm]{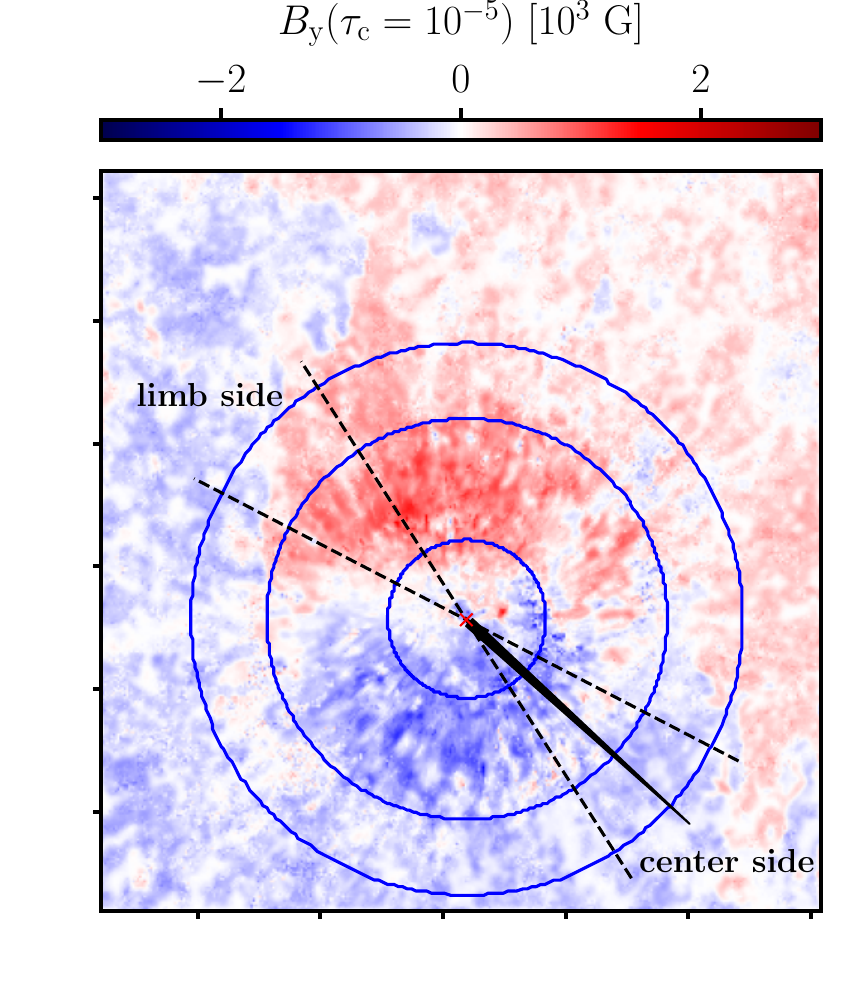}\\
\includegraphics[width=5.8cm]{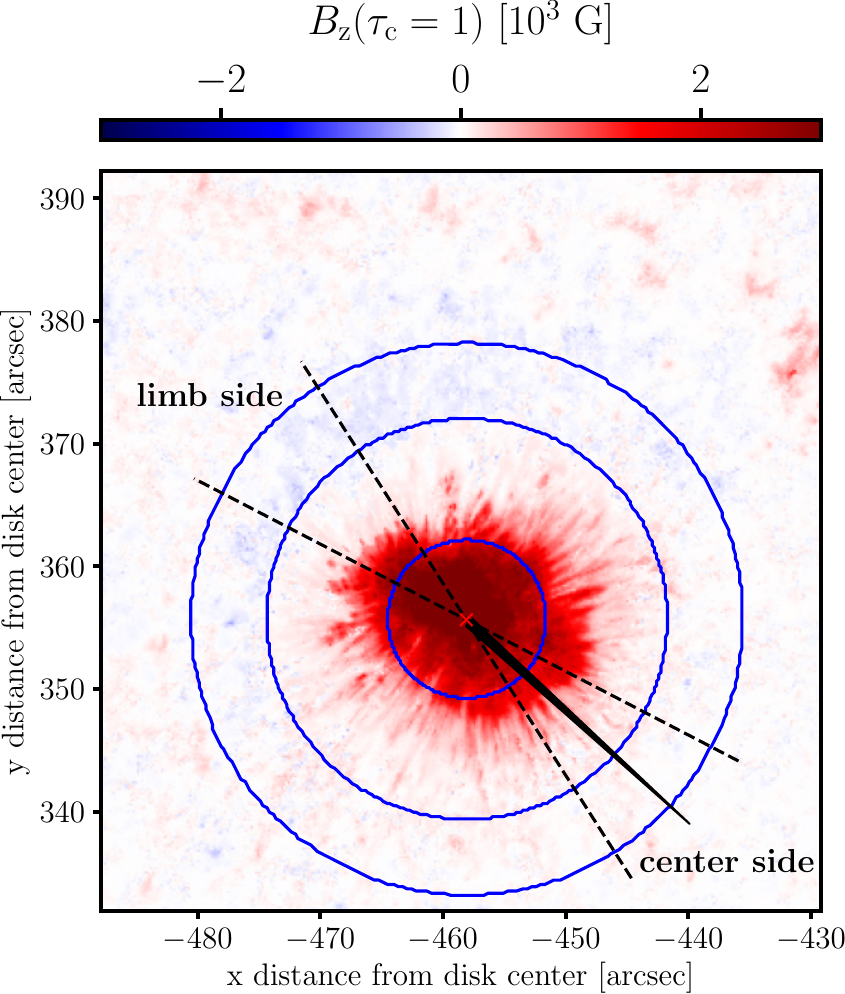} &
\includegraphics[width=5.8cm]{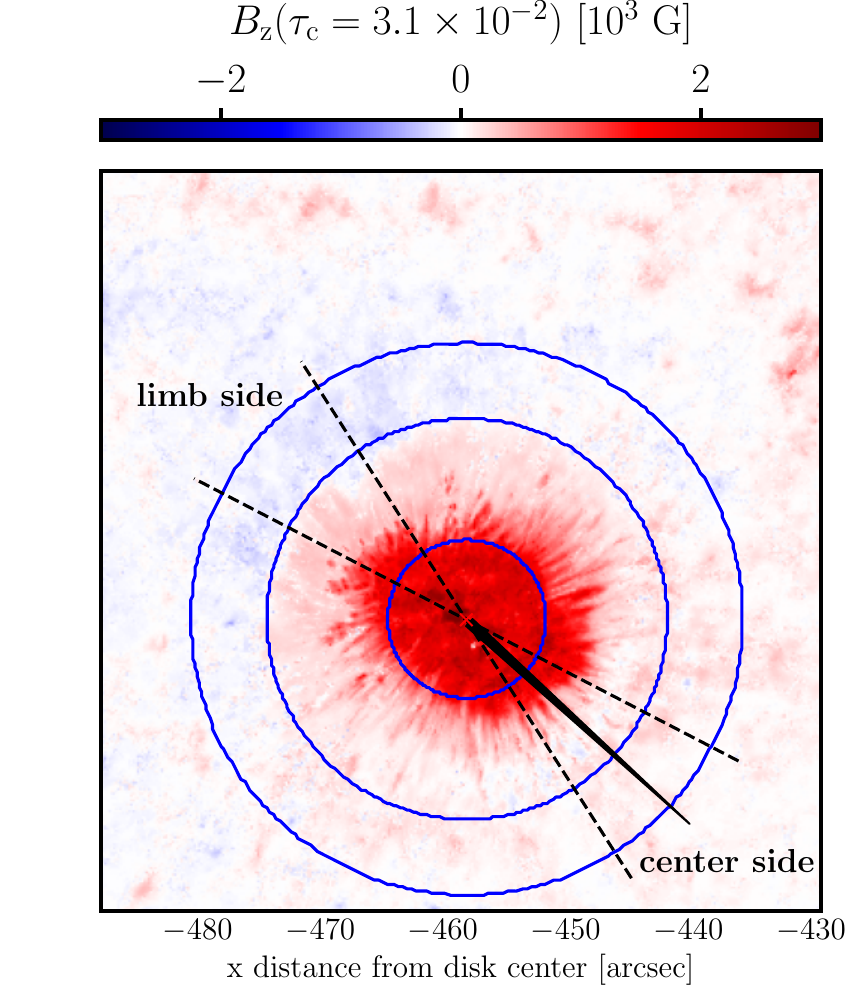} &
\includegraphics[width=5.8cm]{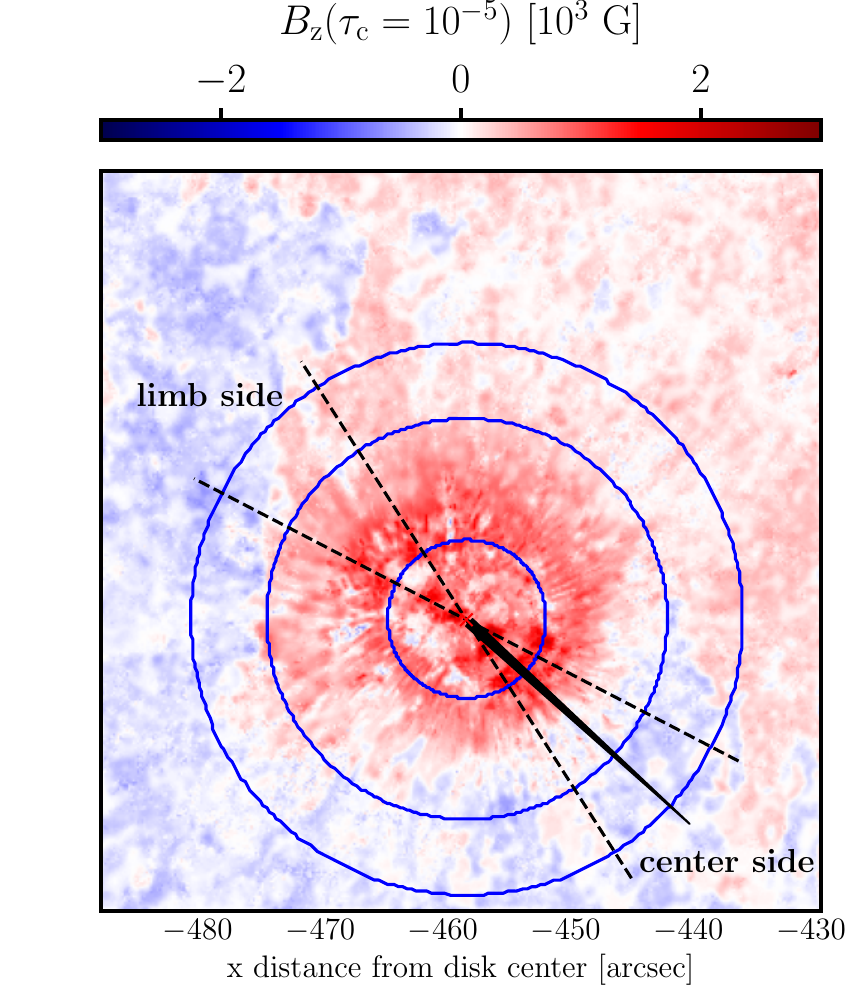}\\
\end{tabular}
\begin{center}
\caption{Inversion results. Similar to middle and bottom panels of Fig.~\ref{fig:results_tau_1} but displaying the $B_x$ (top), $B_y$ (middle) and $B_z$ (bottom) components of the magnetic field in the local reference frame at three different optical depths.\label{fig:results_tau_2}}
\end{center}
\end{figure*}

\subsubsection{Line-of-sight velocity}
\label{subsubsec:vlos}

Velocities in the deep photosphere ($\tau_{\rm c}=1$; bottom-left panel in Fig.~\ref{fig:results_tau_1}) show the common pattern characterized by $v_{\rm los} < 0$ (i.e. velocities towards the observer) in the penumbral region facing the center of the solar disk, and $v_{\rm los} > 0$ (i.e. velocities away from the observer) in the penumbral region facing the solar limb.
This is seen not only in the penumbra, $r/R_{\rm s} \in [0.4,1.0]$, but also beyond the limit of the sunspot's outer boundary, $r/R_{\rm s} > 1$.
While within the sunspot this corresponds to the well known Evershed outflow \citep{evershed1909}, outside it is indicative of the so-called moat outflow \citep{pardon1979moat,nye1988moat}.
In the upper photosphere ($\tau_{\rm c}=10^{-2.5}$; bottom-middle panel) the velocity pattern within the penumbra already switches to the inverse Evershed effect (i.e. inflow towards the sunspot's center).
This supports previous works where the inverse Evershed flow does not appear only as a purely chromospheric phenomena but rather it is already visible in the photosphere \citep{deming1988mg,borrero2008pen}.
The exact optical depth at which the Evershed flow turns into the inverse Evershed flow appears to be located somewhere between $\tau_{\rm c} \in [10^{-2}, 10^{-3}]$ (see also Section~\ref{sec:properties}).\\

Interestingly, while the penumbral Evershed flow reverses into an inflow in the upper photosphere ($\tau_{\rm c} = 10^{-3}$), beyond the visible limit of the spot ($r/R_{\rm s}$) the moat flow continues as an outflow at similar heights. These results suggest that the moat flow is not the continuation of the Evershed flow \citep{johannes2013moat}, otherwise the moat flow should have reversed as well \citep[cf. ]{vargas2007moat}.\\

In the chromosphere ($\tau_{\rm c}=10^{-5}$; bottom-right panel), the inverse Evershed effect completely takes over not only in the penumbra, but also in the moat immediately outside the sunspot. The velocity pattern closely resembles what is usually identified as superpenumbral fibrils. The fact that the moat outflow has now dissipated could be interpreted as the moat outflow being confined to underneath the sunspot's magnetic canopy.\\

In the quiet Sun surrounding the sunspot and beyond the moat $r/R_{\rm s} > 1.4$ there is a pattern of upflows and downflows at $\tau_{\rm c}=1$ (bottom-left panel in Fig.~\ref{fig:results_tau_1}) that does not entirely resemble the well known convective pattern of bright/dark structures being correlated with up/downflows. This happens because our observations are located far from the disk center (see Sect.~\ref{sec:obs}) and therefore granular/intergranular intensity patterns and the inferred line-of-sight velocities appear distorted. This is evidenced by the fact that the line-of-sight velocities in the quiet Sun show clear structuring on the $(x,y)$ plane along the direction perpendicular to the line-of-sight (thick arrow in Fig.~\ref{fig:results_tau_1}). The convective pattern is however, not entirely lost. In fact, the correlation between the observed continuum intensity (upper-left panel) and the inferred $v_{\rm los}(\tau_{\rm c}=1)$ (bottom-left panel) is $-0.23$, lower than the typical values of $-0.6$ at disk center, but still negative, meaning that bright structures show preferentially upflows, whereas dark structures tend to harbor downflows.

\subsubsection{Magnetic field vector}
\label{subsubsec:magneticfield}

Stokes inversion codes, such as FIRTEZ, provide the magnetic field vector $\ve{B}$ in a reference frame where the $z$-axis is aligned with the observer's line-of-sight. This is the so-called observer's reference frame. However, since FIRTEZ corrects the 180$^{\circ}$-ambiguity of the magnetic field in the plane perpendicular to the observer, it is possible to express the magnetic field vector $\ve{B}$ in a new reference frame where the $x$ and $y$ axis are parallel to the equatorial and meridional lines and where the $z$-axis is perpendicular to the solar surface \citep{gary1990lrf}. This is referred to as the local reference frame.\\

Figure~\ref{fig:results_tau_2} presents the maps of these three components of the magnetic field in local reference frame at three different heights in the solar atmosphere.
Owing to the weak observed polarization signals in the Na {\sc i}, Mg {\sc i} and Ca {\sc ii} lines in the quiet Sun, the magnetic field inferred from the inversion in this region is particularly unreliable.
A clear example are the large values of $B_x$, $B_y$
seen at $\tau_{\rm c}=10^{-5}$ (i.e., chromosphere) in the quiet region around the sunspot.
A more detailed description of the errors of the inversion in the observer's reference frame is provided in Table \ref{tab:inversion_errors_regions}.
We note that in the local reference frame, the horizontal and vertical errors have to be computed through error propagation, and therefore will probably lie between the ones shown for $B_z$ and $B_{hor}$ .
Because of all this, we will not discuss the magnetic field inferred in this region.\\

Within the sunspot, Fig.~\ref{fig:results_tau_2} shows that the component of the magnetic field perpendicular to the solar surface $B_z$ is largest at the sunspot's center, while $B_x$ and $B_y$ are largest in the penumbra. In all three cases, the magnetic field components decrease with height, that is, $B_x$, $B_y$ and $B_z$ are larger in the low photosphere ($\tau_{\rm c}=1$) than in the high photosphere and Chrosmosphere ($\tau_{\rm c}=10^{-2.5}, 10^{-5}$, respectively).

\begin{table}[htbp]
\centering
\caption{Mean Inversion Errors in 10x10 pix regions}
\label{tab:inversion_errors_regions}
\begin{tabular}{ll ccc}
\hline \hline
Parameter & Region & $\log \tau = 0$ & $\log \tau = -2.5$ & $\log \tau = -5$ \\
\hline
T [K] & Quiet Sun & 5 & 7 & 58 \\
 & Penumbra & 6 & 10 & 48 \\
 & Umbra & 16 & 50 & 39 \\
\hline
$B_{\mathrm{hor}}$ [G] \tablefootmark{a} & Quiet Sun & 957 & 1008 & 872 \\
 & Penumbra & 68 & 136 & 263 \\
 & Umbra & 395 & 446 & 471 \\
\hline
$B_z$ [G] \tablefootmark{a} & Quiet Sun & 43 & 62 & 92 \\
 & Penumbra & 43 & 71 & 124 \\
 & Umbra & 263 & 369 & 427 \\
\hline
$v_{\mathrm{los}}$ [km/s] & Quiet Sun & 0.26 & 0.14 & 0.19 \\
 & Penumbra & 0.27 & 0.20 & 0.21 \\
 & Umbra & 0.55 & 0.64 & 0.89 \\
\hline
\end{tabular}
\tablefoot{ The errors are computed at constant optical depth.\\
\tablefoottext{a}{Magnetic field is in the observer's reference frame.}
}
\end{table}

\section{Azimuthal averaged sunspot properties}
\label{sec:properties}

\begin{figure*}[ht!]
\label{fig:radial_averages}
\begin{tabular}{ccc}
\hspace{-1cm}
\includegraphics[width=0.36\textwidth]{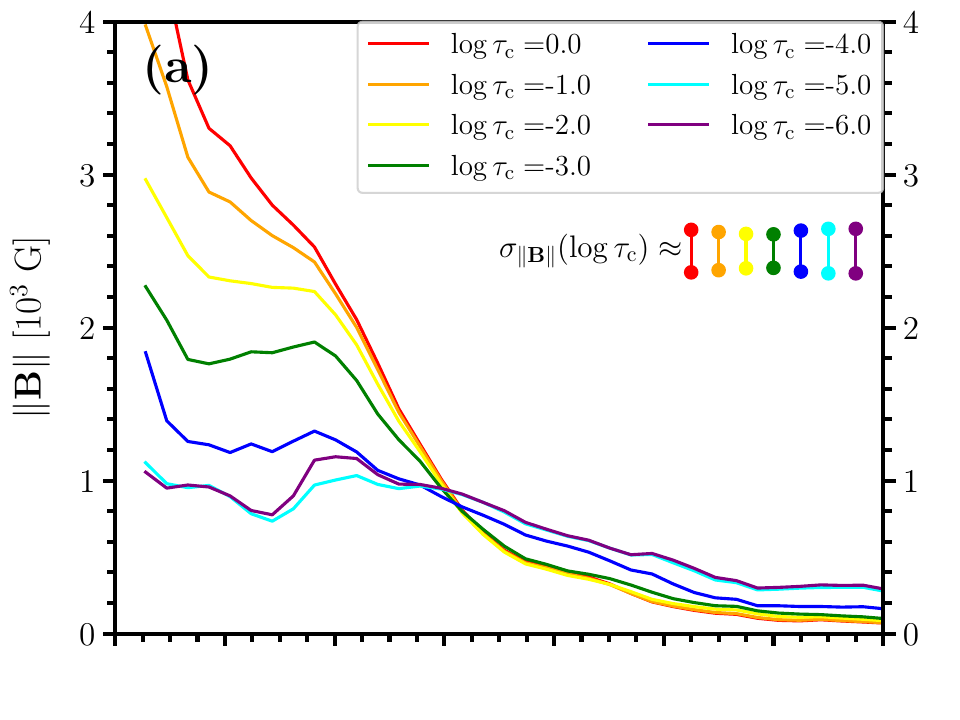} &
\hspace{-0.75cm}
\includegraphics[width=0.36\textwidth]{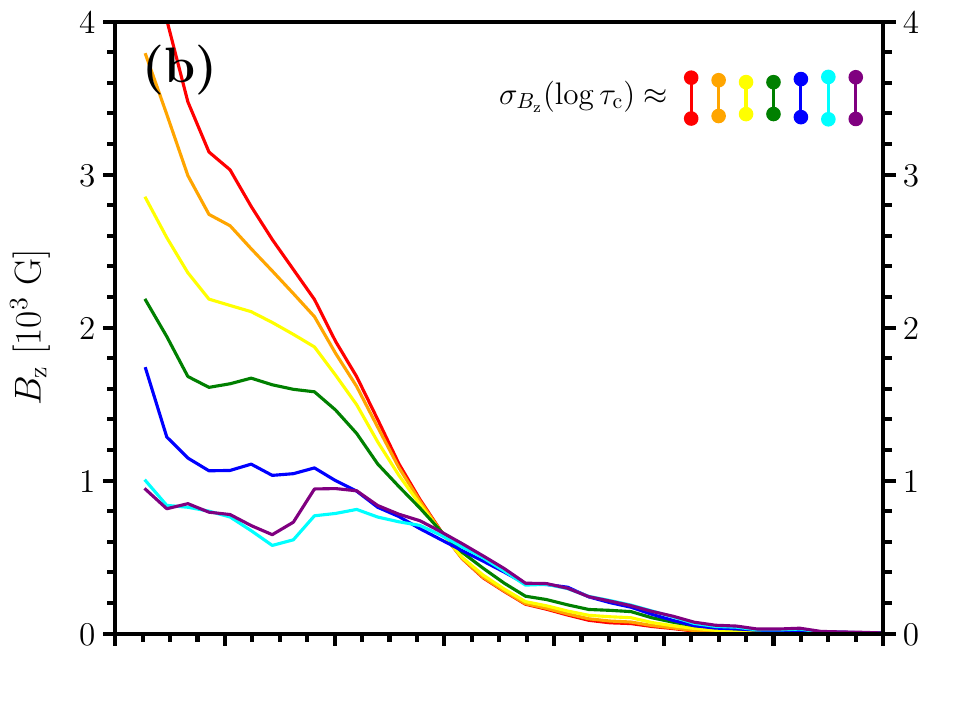} &
\hspace{-0.75cm}
\includegraphics[width=0.36\textwidth]{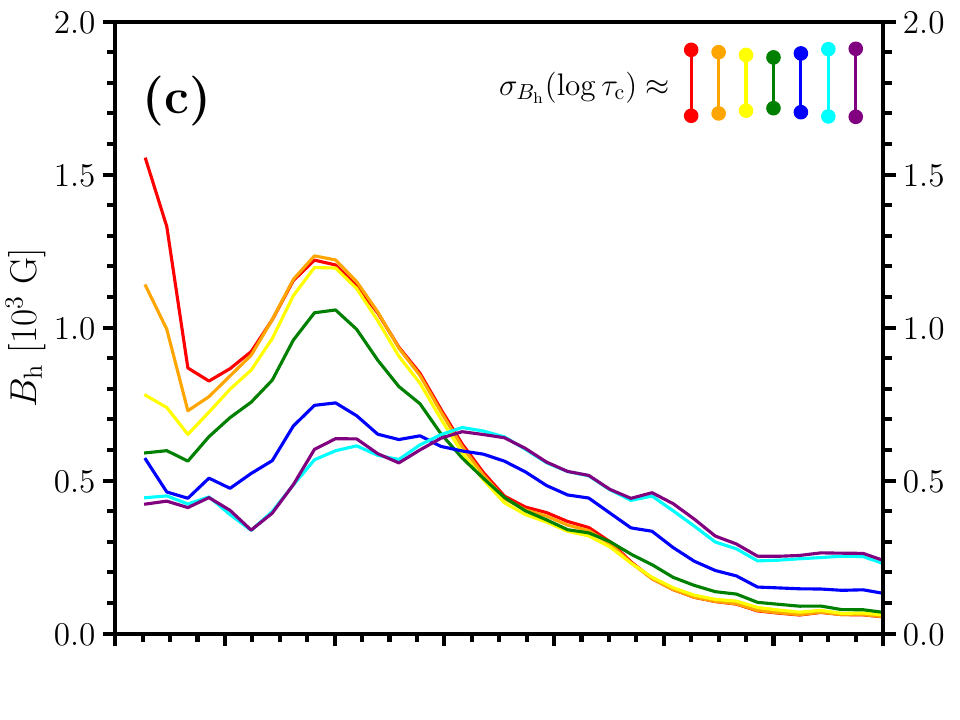} \vspace{-0.35cm} \\
\hspace{-1cm}
\includegraphics[width=0.36\textwidth]{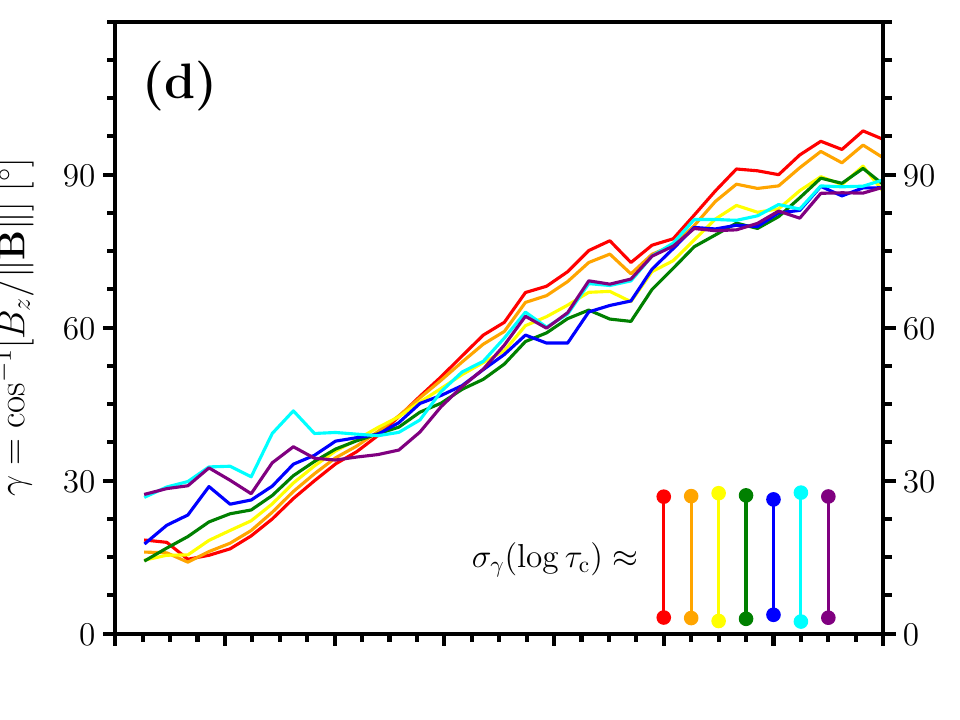} &
\hspace{-0.75cm}
\includegraphics[width=0.36\textwidth]{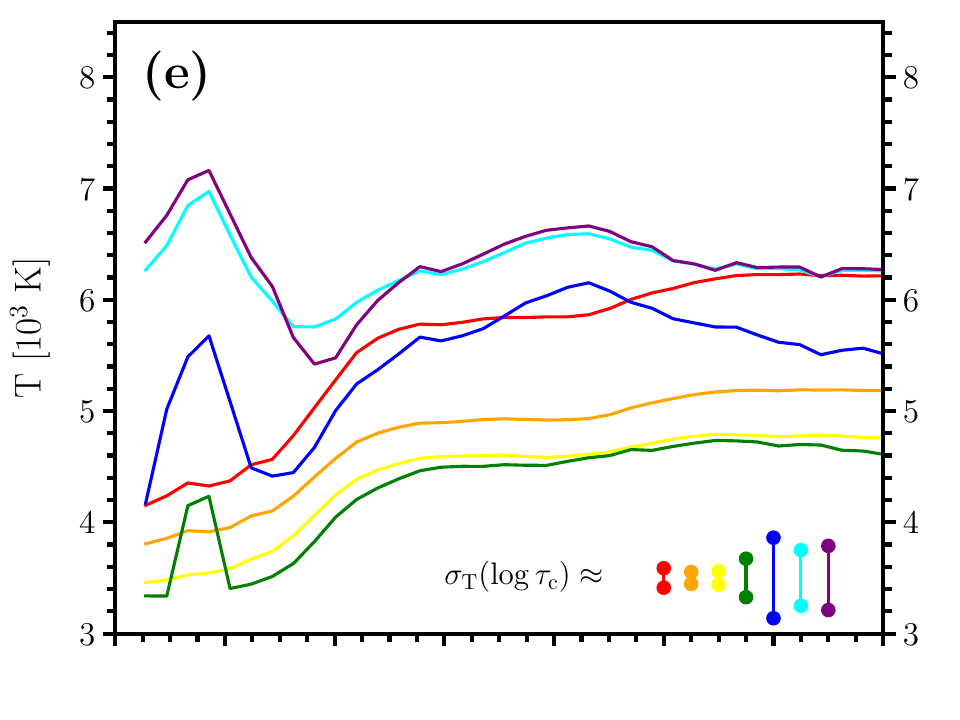} &
\hspace{-0.75cm}
\includegraphics[width=0.36\textwidth]{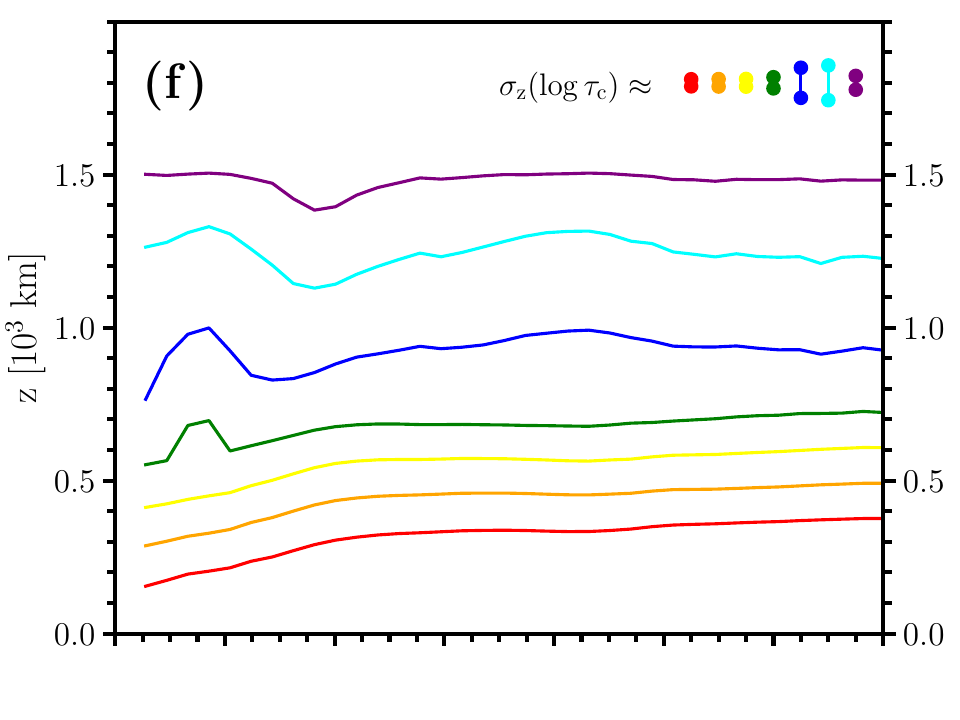} \vspace{-0.35cm} \\
\hspace{-1cm}
\includegraphics[width=0.36\textwidth]{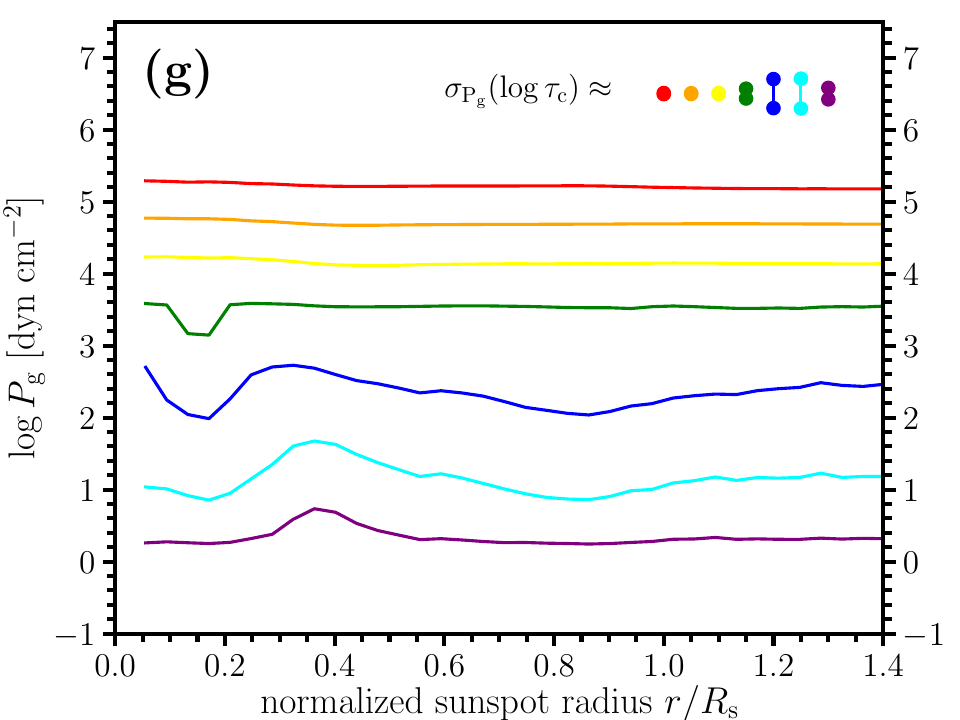} &
\hspace{-0.75cm}
\includegraphics[width=0.36\textwidth]{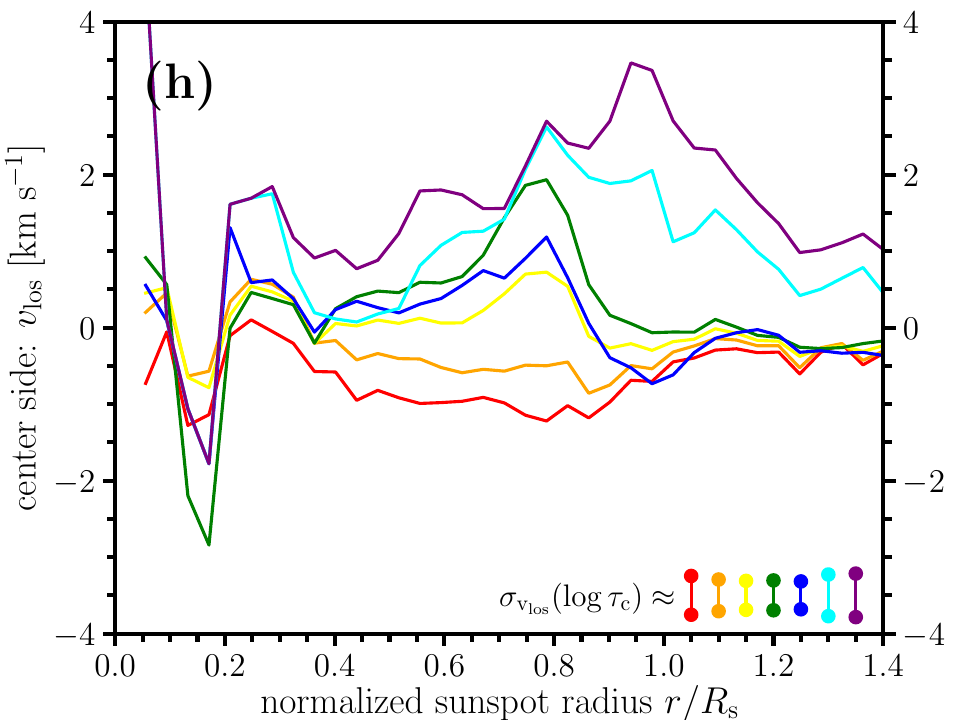} &
\hspace{-0.75cm}
\includegraphics[width=0.36\textwidth]{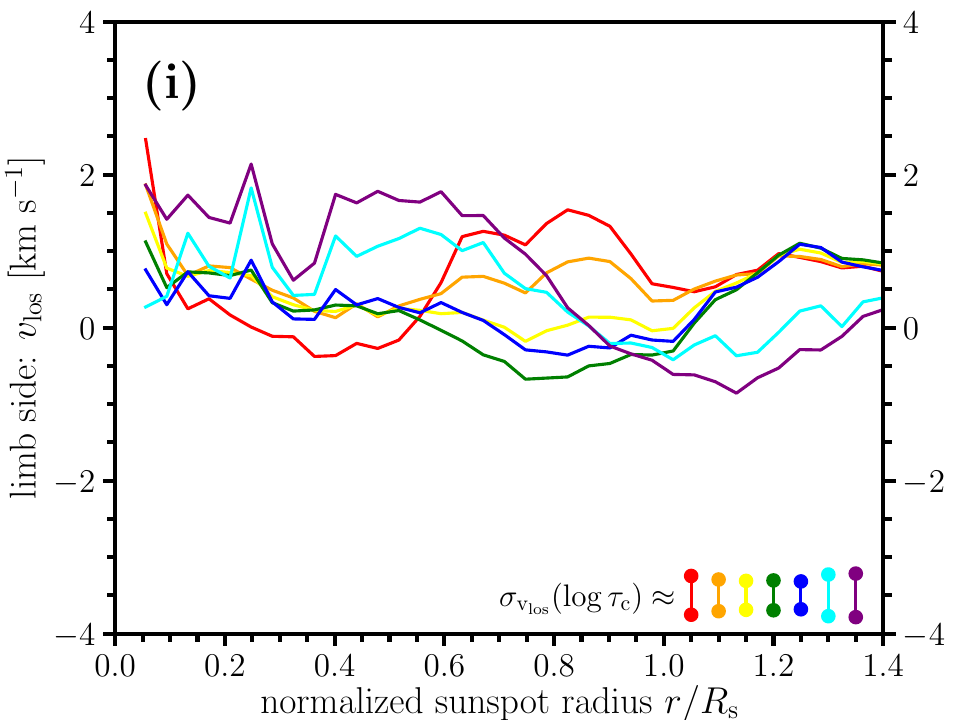}
\end{tabular}
\caption{Average sunspot properties. Azimuthal averages of the physical parameters, as inferred from the Stokes inversion, as a function of the normalized sunspot radius $r/R_{\rm s}$ at different optical-depth levels: $\log\tau_{\rm c} = 0$ (red), $-1$ (orange), $-2$ (yellow), $-3$ (green), $-4$ (blue), $-5$ (cyan), $-6$ (purple). Typical deviations around the mean are represented by the vertical color bars. All panels, expect (h) and (i) were produced by averaging over $2\pi$ radians. Panels h ($v_{\rm los}$ in the center side penumbra) and i ($v_{\rm los}$ in the limb side penumbra) were obtained by averaging only over $\pm \pi/12$ radians around the sunspot's line of symmetry (see cone denoted by dashed lines in Figs.~\ref{fig:results_tau_1} and ~\ref{fig:results_tau_2}).\label{fig:rz}}
\end{figure*}

In this section, we present the azimuthal averaged physical properties of the observed sunspot as a function of normalized radial distance from the sunspot's center at different optical depth levels $(r/R_{\rm s},\log\tau_{\rm c})$. We note that, the very center of the umbra is avoided ($r/R_{\rm s} \ge 0.05$) because at short radial distances we have very few points to average over. Results are presented in Fig.~\ref{fig:rz} and include: module of the magnetic field vector ($\|\ve{B}\|$; panel-a), vertical component of the magnetic field in the local reference frame ($B_{\rm z}$; panel-b), horizontal component of the magnetic field in the local reference frame ($B_{\rm h}$; panel-c), inclination of the magnetic field with respect to the direction normal to the solar surface ($\gamma$; panel-d), temperature ($T$; panel-e), Wilson depression ($z$; panel-f), gas pressure ($P_{\rm g}$; panel-g), line-of-sight velocity in the center side of the penumbra ($v_{\rm los}$; panel-h), and line-of-sight velocity in the limb side of the penumbra ($v_{\rm los}$; panel-i). We note that since high spatial resolution is not needed in order to perform spatial averages, we do not use the results of the Stokes inversion of the diffraction limited data, but rather the results of the inversion of the spatially binned data (see Sect.~\ref{sec:obs}).\\

With the exception of $v_{\rm los}$, all azimuthal averages are performed over $2\pi$ angles around concentric circles for increasing radial distances around the sunspot's center (see examples blue circles in Figs.~\ref{fig:results_tau_1} and \ref{fig:results_tau_2}). In the case of the line-of-sight velocity we do not perform $2\pi$ averages because, due to the photospheric Evershed flow redshifted velocities on the limb side of the penumbra will cancel with blue shifted velocities on the center side (see lower panels in Fig.~\ref{fig:results_tau_1}). Likewise, due to the inverse Evershed low in the chromosphere, redshifted velocities in the center side of the penumbra will cancel with blue shifted velocities on the center side. Instead of this, the averaged line-of-sight velocities are calculated in a cone of $\pm 15^{\circ}$ around the line-of-symmetry of the sunspot (see dashed lines in Figs.~\ref{fig:results_tau_1} and \ref{fig:results_tau_2}).\\

In adittion to the mean values of the physical parameters as a function of $(r/R_{\rm s},\log\tau_{\rm c})$ we also display, as vertical bars in Fig.~\ref{fig:rz}, typical standard deviations of these physical parameters at different optical depths. We note that these bars do not represent errors in the inversion, but rather standard deviations around the mean value. In this regard, it is important to note that parameters such as $\|\ve{B}\|$, $B_{\rm z}$, $B_{\rm h}$ and $\gamma$ (panels a,b,d and d, respectively) feature a large deviation around the mean because of the presence of the penumbral fine structure, where regions of strong and vertical magnetic field (i.e. spines) are interlaced azimuthally with regions of much weaker and horizontal magnetic field \citep[i.e. intraspines;][]{borrero2004pen,borrero2025pen}.\\

The results presented in Fig.~\ref{fig:rz} are unique and expand over previous works \citep{title1993,carlosw2001,mathew2003pen,luis2004pen,cuberes2005,borrero2011review} in that we extend our analysis to part of the moat ($r/R_{\rm s}=1.4$) and to the temperature minimum and lower chromosphere ($\log\tau_{\rm c} \in [-4,-6]$). Moreover, because our inversions were carried out using a three-dimensional magneto-hydrostatic equilibrium instead of vertical hydrostatic equilibrium (see Sect.~\ref{subsec:inversion}) we can provide more reliable values for the gas pressure and Wilson depression (see panels h and g in Fig.~\ref{fig:rz}) than those reported in previous works. Based on the various panels in this figure we can draw a number of additional observations:

\begin{enumerate}
\item in the umbra and mid-inner penumbra ($r/R_{\rm s} < 0.6$) we find that $\textrm{d}B/\textrm{d}\tau_{\rm c} >0$, whereas in the outer penumbra  ($r/R_{\rm s} \ge 0.6$) this derivative changes sign: $\textrm{d}B/\textrm{d}\tau_{\rm c} <0$. This is reminiscent of previous findings by \citet[][see their Fig.~4]{borrero2008penb} even though that work focused mostly on the penumbral intraspines, whereas here we have azimuthally averaged over both spines and intraspines.

\item in the moat and superpenumbra ($r/R_{\rm s} \ge 1.0$) the trend $\textrm{d}B/\textrm{d}\tau_{\rm c} <0$ continues. Moreover, the radial component of the magnetic field $B_{\rm h}$ (panel c) in the moat/super-penumbra is larger in the chromosphere than in the photosphere. Although it might be tempting to interpret this result in terms of the presence of a magnetic canopy, we refrain ourselves from such interpretation as we have already established that the inference of the magnetic field outside the spot at chromospheric heights is not reliable due to the low polarization signals observed in the Ca {\sc ii} spectral line.

\item physical parameters such as the temperature (panel e), Wilson depression (panel f), line-of-sight velocity on the center side of the penumbra (panel h) feature large variations (seen sometimes as dips, sometimes as peaks) very close to the umbral center ($r/R_{\rm s} \approx 0.1$). These are produced by the presence of an umbral flash very close to the umbral center. This particular event will be studied in detail in Sect.~\ref{sec:flash}.

\item on average the Wilson depression, $z(\tau_{\rm c}=1)$, is located about 250~km deeper inside the umbra than in the quiet Sun (see panel f). Although this might be seen as too shallow, we emphasize that the largest value for the Wilson depression occurs at the very umbral center (which was avoided in these figures) and amounts to about 350~km.

\item current values of the gas pressure $P_{\rm g}$ at different optical depth levels (panel g) look rather constant with radial distance. This is an effect of having used iso-surfaces of optical depth. At fixed geometrical heights $z$ the gas pressure would show much stronger variations with radial distance.

\item according to panels h and i, in the penumbra $r/R_{\rm s} \in [0.4,1.0]$, the regular Evershed flow is still seen at optical depths $\tau_{\rm c} = 10^{-2}$ (see yellow lines), whereas for optical depths $\tau_{\rm c} = 10^{-3}$ (green lines) we already detect the inverse Evershed flow.

\end{enumerate}

\section{Instantaneous view of an umbral flash}
\label{sec:flash}

\begin{figure*}[ht!]
    \vspace{-3.5cm}
    \hspace{-2cm}
    \includegraphics[width=1.3\linewidth]{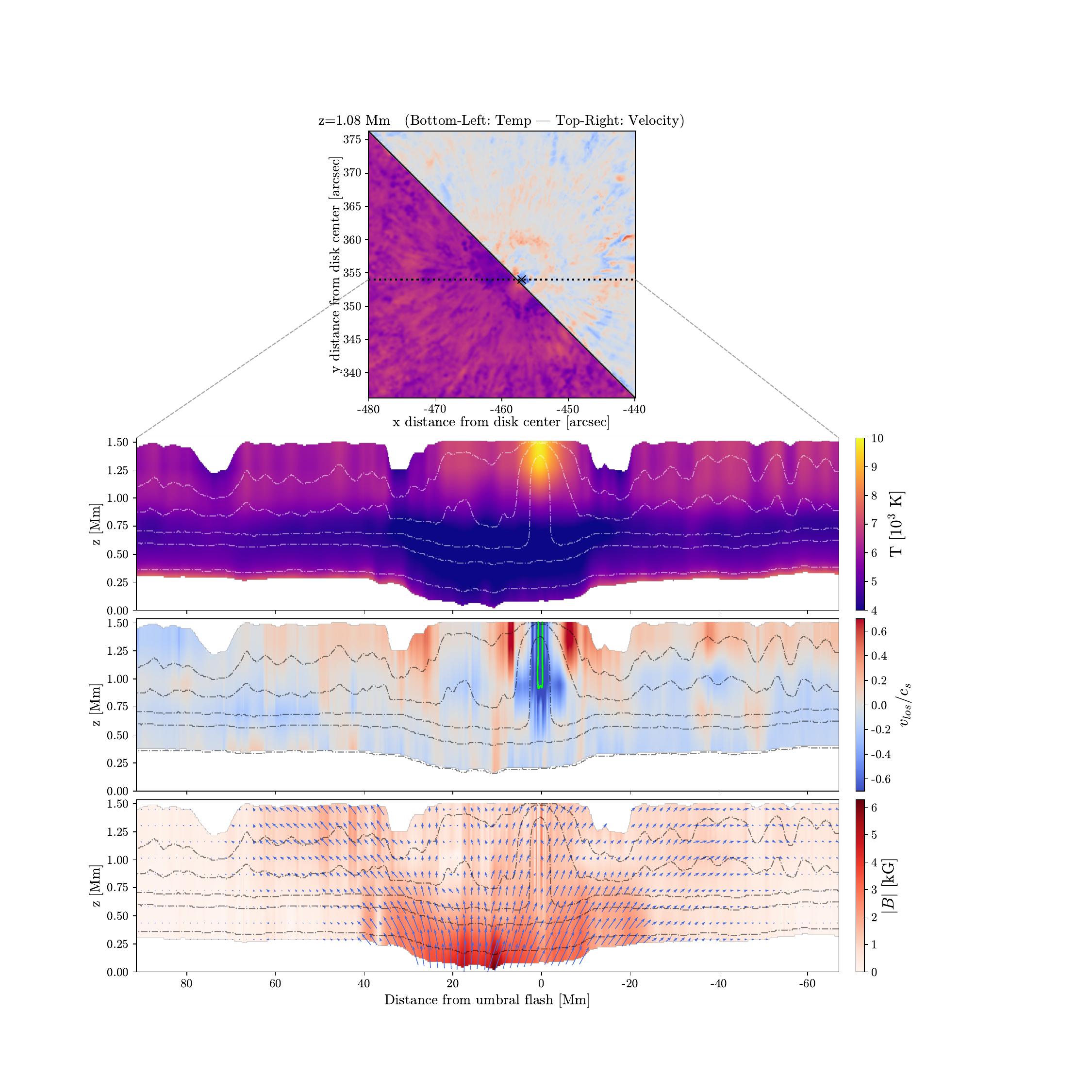}
    \vspace{-2cm}
    \caption{View of the umbral flash inferred from the diffraction limited data. Upper panel: divided map of temperature $T(x,y)$ (bottom left) and line-sight-velocity $v_{\rm los}$ (top right) at $z \sim 1$~Mm. The dotted horizontal line represents the slice selected for the
    three lower panels, and the cross indicates where the umbral flash happens (i.e. where the strongest supersonic velocities are found). Bottom panels: temperature $T(x,z)$, Mach number $M(x,z)$, and magnetic field ${\bf B}(x,z)$
    on the plane of the slice indicated on the upper panel ($y=88$~Mm). As a reference, the curves of constant optical depth $\log\tau_{\rm c} = 0, -2, -3, -4, -5$ are indicated by the black or white dashed lines. Green
    contour indicates the regions where supersonic flows are found: $\|M > 1\|$.}
    \label{fig:temperature_velocity_umbral_flash_cut}
\end{figure*}

As already mentioned in Sect.~\ref{sec:properties}, an important aspect to notice in this particular dataset is the presence of an umbral flash or small-scale umbral brightening. Umbral flashes are known to be transient events that occur with a periodicity of about 3 minutes \citep{hector2000uf} and unfortunately the analyzed data can only provide a picture of this phenomena at an single instant in time. The umbral flash in our data occurs very close to the umbral center at $(x,y) \approx (-458\arcsec, 352\arcsec)$. In the spatially averaged data (Fig.~\ref{fig:results_tau_1}) there is a sudden temperature enhancement at this location in the high photosphere ($\tau_{\rm c} = 10^{-2.5}$). The enhancement spreads over a larger horizontal area at $\tau_{\rm c} = 10^{-5}$ (i.e. chromosphere). These temperature enhancements are produced by the presence of the aforementioned umbral flash \citep{beckers1969flash,wittmann1969flash}. Remarkably, while the core of the Ca {\sc ii} spectral line responds by increasing its intensity (see upper-right panel in Fig.~\ref{fig:results_tau_1}) and turning into emission (see red lines in Fig.~\ref{fig:sample_fits}), the core of the Mg {\sc i} line does not show any increase in the observed intensity (see upper-middle panel in Fig.~\ref{fig:results_tau_1}) in response to the aforementioned temperature increase. The core of the Na {\sc i} line behaves similarly to the core of Mg {\sc i} and displays no intensity enhancement (not shown). The reason for this different behavior is provided in Appendix~\ref{app:betas}, where we show that the Na {\sc i} and Mg {\sc i} lines strongly decouple from the local temperature in the upper photosphere and chromosphere, whereas the Ca {\sc ii} line stays more strongly coupled to the temperature.\\

In order to study the umbral flash in more detail we turn to the results of the inversion of the diffraction limited data (see Sect.~\ref{sec:obs}). Results for the temperature $T(x,y)$ and line-of-sight velocity $v_{\rm los}(x,y)$ at a constant geometrical height of $z=1.08$~Mm, obtained from the Stokes inversion of the diffraction limited data, are presented in Fig.~\ref{fig:temperature_velocity_umbral_flash_cut} (upper panel).
This figure also displays, in the bottom panels, the temperature $T$, line-of-sight velocity normalized to the local sound speed or Mach number $M = v_{\rm los}/c_{\rm s}$, and magnetic field $\mathbf{B}(y,z)$
in a $(x,z)$ plane for an $y$ value that cuts horizontally through the umbral flash (see dashed line).
The local sound speed has been calculated assuming an adiabatic constant of $\gamma= 5/3$ (i.e. monoatomic gas).\\

At the location of the umbral flash and its surroundings, there is a region where large temperature enhancements are found in the upper photosphere and chromosphere ($\tau_{\rm c} \in [10^{-3},10^{-6}]$).
At these locations supersonic upflow velocities ($M < -1$) are also detected.
These regions are marked within green contours, in the third row panel in Fig.~\ref{fig:temperature_velocity_umbral_flash_cut}.
The magnetic field displayed in the bottom most panel is converted to the local reference frame and displays typical sunspot behavior. Where the flux tube is mostly vertical in the center (of around $\sim 5000$G) and opens towards the penumbra, becoming more horizontal.
The umbral flash regions show disturbances both in the strength ($\Delta |B| \sim 1$KG) and direction, changing mostly the horizontal components.
Magnetic field variations show both weakening and enhancements, consistent with \cite{vasco2017}.
Values as high as $\| M \| \geq 1.5$ are sometimes inferred.
We note that the previous inequalities on the supersonic velocities are because they are computed from the line-of-sight component of the velocity alone, and therefore, even larger values are to be expected if we consider the full velocity vector.
Unfortunately, measurements of the components of the velocity vector in the plane perpendicular to the line-of-sight are usually not possible \citep[cf.][]{helena2026}.
The existence of these supersonic velocities bespeaks the presence of a shock front \citep{jaime2013sunspot}.
In agreement with previous studies \citep{vasco2017, vasco2019, vasco2020uf} we also find that the supersonic upflow is surrounded by large, albeit still subsonic, downflowing velocities.
\cite{vasco2020uf}[see Sect.~4.2] have argued that there is an intrinsic degeneracy in the inference of the temperature in umbral flashes by Stokes inversion codes.
This degeneracy arises because an increase in the source function needed to produce emission in the line core of the Ca {\sc ii} line (see red lines in the upper panel of Fig.~\ref{fig:sample_fits}) can be caused by either an increase in the density or an increase in the temperature.
Now, because Stokes inversion codes have traditionally operated under the assumption of vertical hydrostatic equilibrium, they cannot produce local enhancements in the density.
This is not the case for the FIRTEZ inversion code, where three-dimensional magneto-hydrostatic equilibrium can produce those local enhancements through the action of the Lorentz force.
We note that this is the first time such studies are carried out, and therefore we can better constrain the origin of the source function enhancement in the umbral flash region.

However, according to \cite{vasco2020uf}, the reason the density enhancements occur is due to the presence of converging flows, and these are not accounted for in FIRTEZ.
To do so, FIRTEZ would have to determine the density and gas pressure not under magneto-hydrostatic conditions but rather under magneto-hydrostationary conditions (i.e. including the advective velocity term,  $\rho (\ve{v} \cdot \nabla) \ve{v}$, in the momentum equation).
Adding this term would be particularly beneficial in the case of umbral flashes due to the presence of supersonic velocities.\\

\begin{figure}[ht!]
    \centering
    \includegraphics[width=1\linewidth]{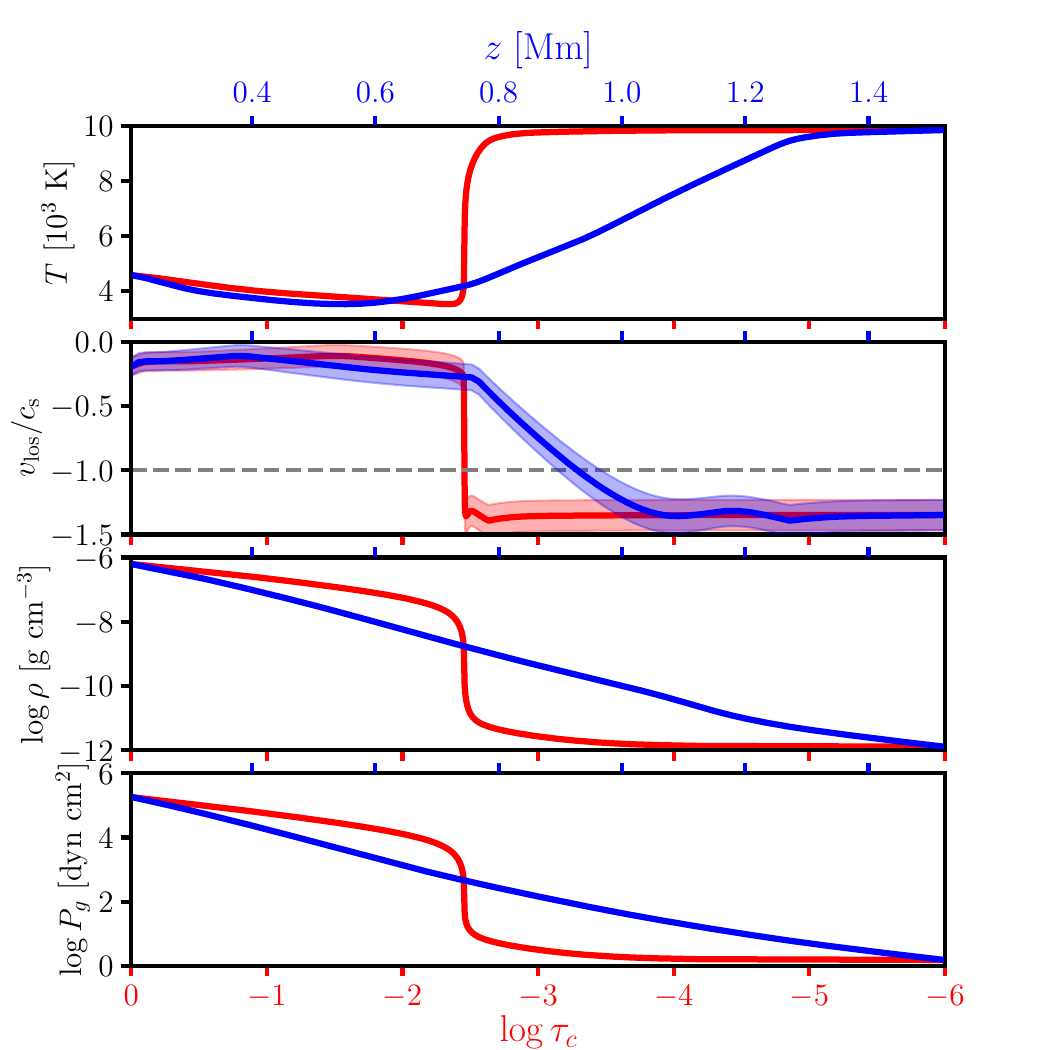}
    \caption{Atmosphere stratification of one of the pixels of the umbral flash with supersonic velocities. From top to bottom: temperature $T$, line-of-sight velocity normalized to the local sound speed (i.e. line-of-sight Mach number) $v_{\rm los}/c_{\rm s}$, logarithm of density $\log\rho$, and logarithm of gas pressure $\log P_{\rm g}$. Bottom axis and red lines represent the physical parameters as a function of the optical depth scale $\log\tau_{\rm c}$, whereas the upper axis and blue lines represent the physical parameters as a function of geometrical height $z$. We note the continuum ($\log\tau_{\rm c} = 0$) is formed at around $z \approx 0.2$~Mm above the lowermost boundary of the domain.
    Error estimations inferred from the Stokes inversion are presented for the temperature and velocity as shadowed areas.
    The horizontal dashed line represents the limit at which the line-of-sight velocity becomes supersonic.}
    \label{fig:shock_profile}
\end{figure}

Figure~\ref{fig:shock_profile} shows the thermodynamic (temperature $T$, gas pressure $P_{\rm g}$, and density $\rho$) as well as kinematic (Mach number $M$) parameters present in one pixel where the aforementioned shocks are detected. These physical parameters are displayed as a function of the geometrical height $z$ (blue lines), and as a function of the logarithm of the continuum optical depth $\log\tau_{\rm c}$ (red lines). As can be seen, there are clear supersonic velocities and temperature enhancements in the upper photosphere and chromosphere: $\log\tau_{\rm c} < -2.5$.
It is worth noticing that the consideration of the supersonic regime of some inverted pixels already takes into account the uncertainties of the inversion process
as described in Sec. \ref{subsec:results}.
Similar results to those shown in Fig.~\ref{fig:shock_profile} also appear in several pixels around the umbral flash.
Temperature errors, although also included in this figure (top panel), are below the line width ($ 20 < \delta_T < 100$ K).\\

Comparison of both vertical scales, $z$ and $\log\tau_{\rm c}$, in combination with the optical depth surfaces (shown as colored dashed lines in Fig.~\ref{fig:temperature_velocity_umbral_flash_cut}) suggests that the opacity greatly increases at the top of the atmosphere, where the temperature enhancement is located, and then drops significantly underneath until the optical depth matches with the surrounding columns at around $\log \tau_{\rm c} \sim -2.5$.
This phenomenon compresses the tau scale in z at the higher layers, removing the sensitivity in the $\tau$ scale between the umbral flash heights and the base of the umbral region.

\section{Conclusions}
\label{sec:conclusions}

In this paper we have analyzed spectropolarimetric data of a sunspot that include multiple spectral lines formed in the photosphere and chromosphere.
The spectral lines were recorded with the CRISP instrument \citep{scharmer2008crisp} attached to the 1-m SST Telescope.
The analysis was performed with the FIRTEZ Stokes inversion code \citep{adur2019firtez} which provides us with the sunspot's kinematic, thermodynamic and magnetic properties.\\

Owing to non Local Thermodynamic Equilibrium effects being included in the analysis, our inferences are valid from the deep photosphere to the mid chromosphere simultaneously, which makes it particularly interesting.
This corresponds to a height range that spans up to six orders of magnitude of optical depth: $\log\tau_{\rm c} \in [0,-5]$.
In addition, the Stokes inversion accounted for magneto-hydrostatic constraints and therefore the inferences are also valid in the geometrical height scale \citep{borrero2021firtez}.
In this case, the range of heights where the physical parameters are $z-z(\tau_{\rm c}=1) \in [0,1]$~Mm, where $z(\tau_{\rm c}=1)$ refers to the layer where the continuum is formed (see e.g. black dashed line in the bottom panels of Fig.~\ref{fig:temperature_velocity_umbral_flash_cut}).\\

With these results we have studied the average sunspot properties (such as temperature, line-of-sight velocity, magnetic field vector, gas pressure, Wilson depression, etc.) as a function of radial distance from the sunspot's center and as a function of geometrical height and optical depth $(r, z, \tau_{\rm c})$.
This work extends similar studies that concerned mostly the photospheric properties \citep{mathew2003pen,borrero2011review} to include also the chromosphere and expand the analysis giving a tomographic view of the sunspot thanks to the coupling capabilities that FIRTEZ inversion code provides.
We have also extended previous analysis to radial distances beyond the penumbra-quiet Sun boundary ($r/R_{\rm s} > 1$) to include part of the moat and superpenumbra.\\

In addition, very close to the umbral center, an umbral flash is detected in our data. Although umbral flashes are characterized by oscillatory behavior in time \citep{hector2000uf,vasco2020uf}, the data we have analyzed here only capture a single time instance and therefore cannot be used to fully describe these phenomena.
However, at the particular time of our observations, the umbral flash appears in the upper photosphere and chromosphere ($\tau_{\rm c} \in [10^{-3},10^{-6}]$) where it manifests itself with large temperature and opacity enhancements. In addition, supersonic upflowing line-of-sight velocities $v_{\rm los}/c_{\rm s} \approx -1.5$ are seen at the center of the umbral flash. These are surrounded by large, subsonic, downflowing line-of-sight velocities $v_{\rm los}/c_{\rm s} \approx 0.5$. These results support the idea that umbral flashes are caused by converging flows that produce hydrodynamic shocks \citep{vasco2020uf}. In the future we will try to analyze a time-series of these observations and fully characterize the temporal behavior of the umbral flash in the correct $z$-scale by means of the non-LTE magnetohydrostatic Stokes inversion described in this paper.

\begin{acknowledgements}
The development of the FIRTEZ inversion code is funded by two grants from the Deutsche Forschung Gemeinschaft (DFG): projects 321818926 and 538773352. AVA acknowledges support from the DFG (project 538773352). APY acknowledges support from the Swedish Research Council (grant 2023-03313). This project has been funded by the European Union through the European Research Council (ERC) under the Horizon Europe program (MAGHEAT, grant agreement 101088184). AP was supported by grant PI~2102/1-1 from the Deutsche Forschungsgemeinschaft (DFG). IK was supported by the Deutsche Forschungsgemeinschaft (DFG) project number KO 6283/2-1. This research data leading to the results obtained has been supported by SOLARNET project that has received funding from the European Union’s Horizon 2020 research and innovation programme under grant agreement no 824135. This research has made use of NASA's Astrophysics Data System and of the atomic data publicly provided by the National Institute of Standards and Technology (US Department of Commerce).
\end{acknowledgements}

\bibliographystyle{aa}
\bibliography{ms}

\clearpage
\appendix

\section{Spectral lines and height sensitivity}
\label{app:response_functions}

\begin{figure}[ht!]
\begin{tabular}{cc}
\includegraphics[width=4cm]{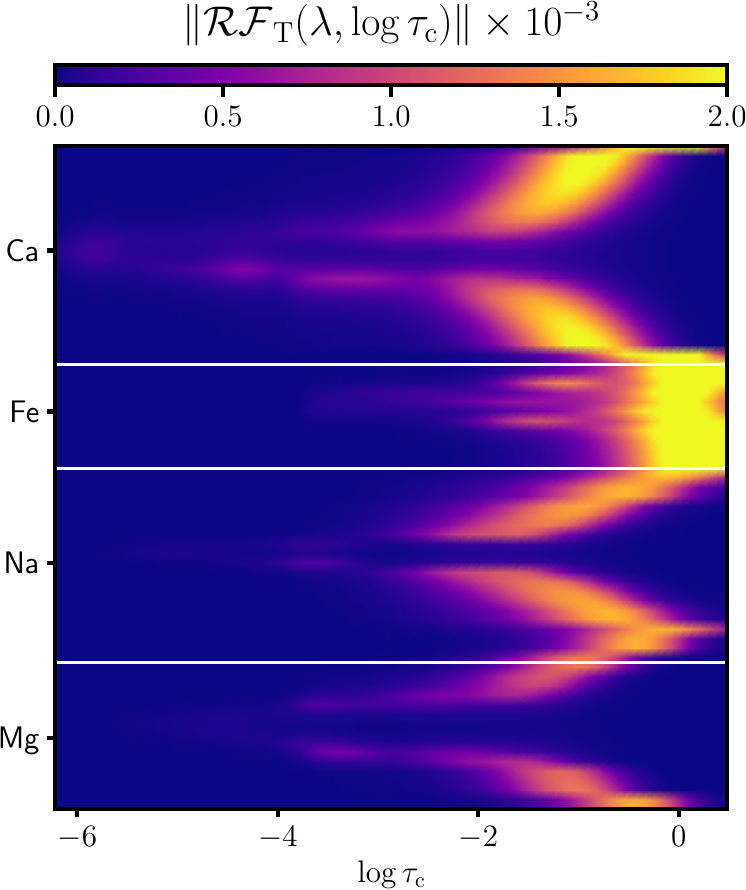} &
\includegraphics[width=4cm]{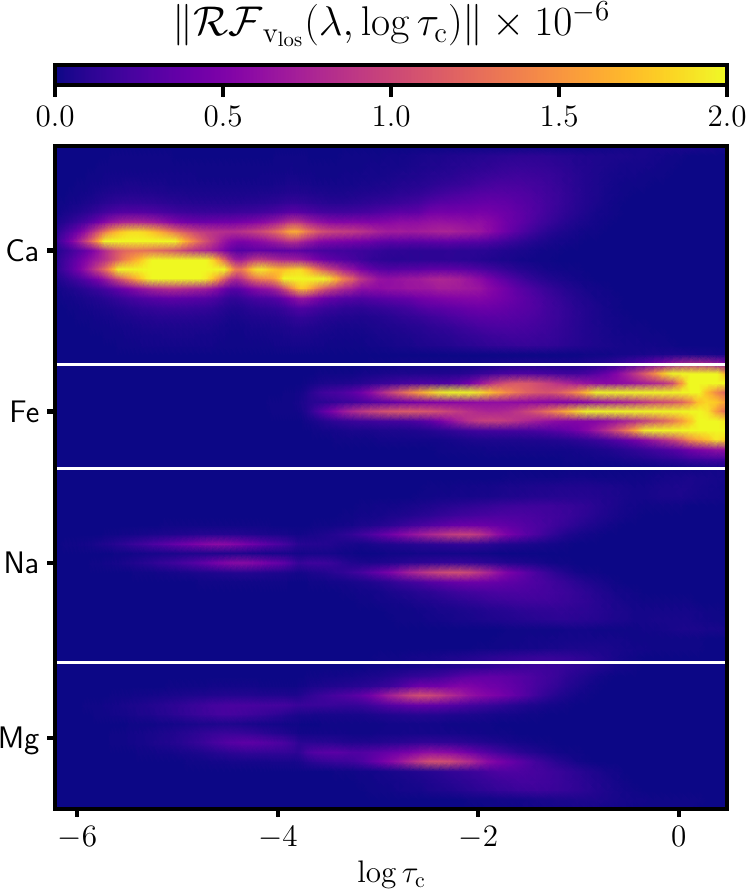} \\
\includegraphics[width=4cm]{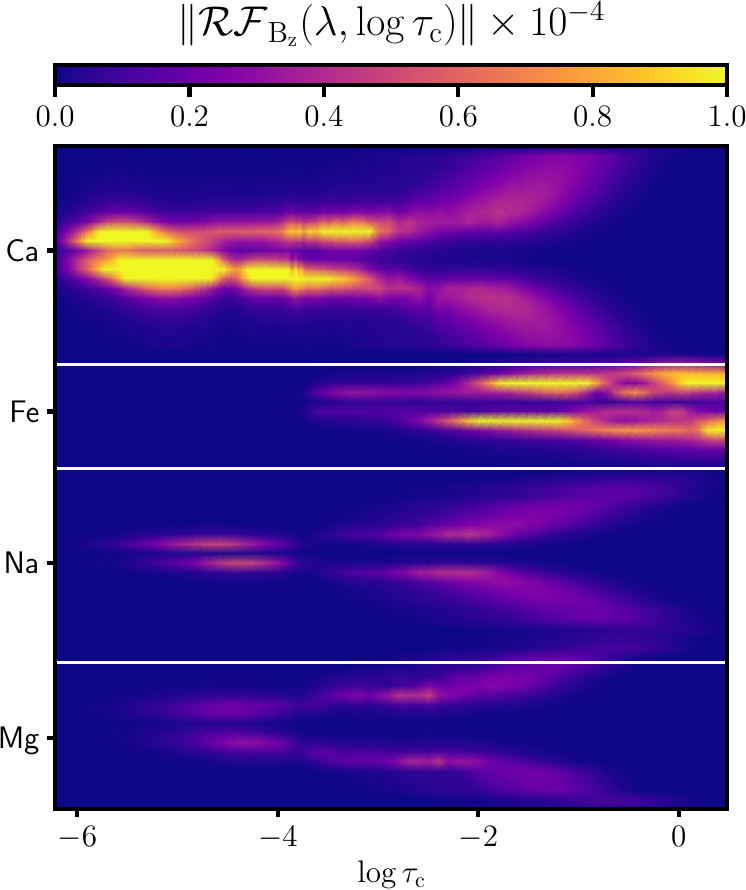} &
\includegraphics[width=4cm]{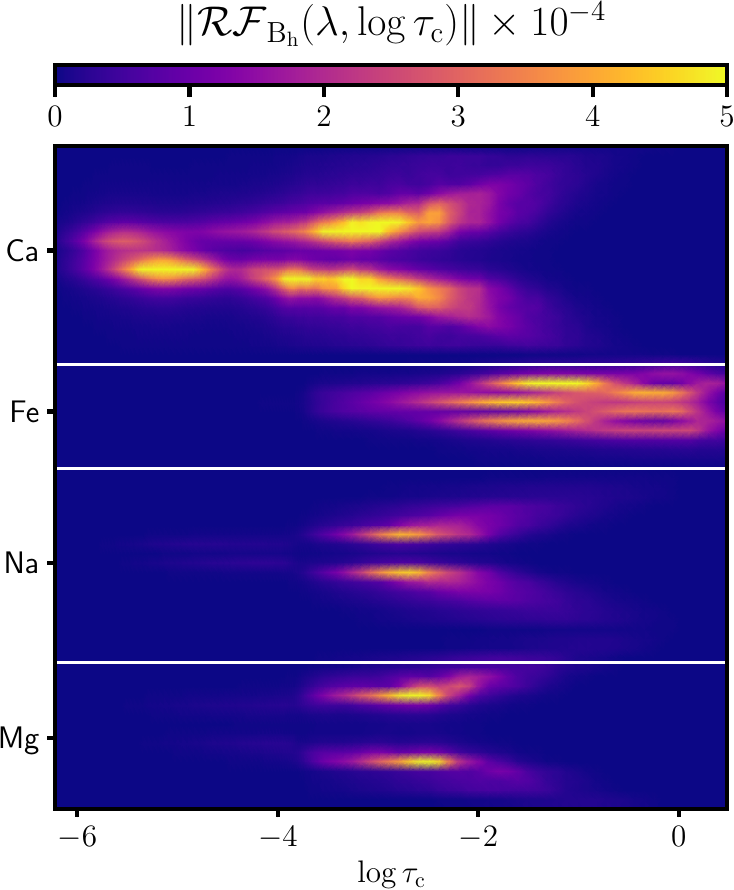}
\end{tabular}
\caption{Response functions for the temperature ($\mathcal{RF}_{\rm T}$; upper-left), line-of-sight velocity ($\mathcal{RF}_{\rm v_{\rm los}}$; upper-right), vertical component of the magnetic field ($\mathcal{RF}_{\rm B_{\rm z}}$; lower-left), and horizontal component of the magnetic field ($\mathcal{RF}_{\rm B_{\rm h}}$; lower-right) for a pixel located in the penumbra as a function of wavelength and optical depth $\log\tau_{\rm c}$. See text for details.\label{fig:rf}}
\end{figure}

In Figure~\ref{fig:rf} we show spatially averaged response functions to several physical parameters for the observed spectral lines. They were calculated by averaging the absolute value of the response functions obtained from the atmospheric models as inferred from the Stokes inversion (see Section~\ref{subsec:inversion}) in a region located in the penumbra and containing 10$\times$10 pixels. All four Stokes parameters $(I,Q,U,V)$ were used to calculate $\mathcal{RF}_{\rm T}$ and $\mathcal{RF}_{\rm v_{\rm los}}$ but only Stokes $V$ to determine $\mathcal{RF}_{\rm B_{\rm z}}$ and only $Q$ and $U$ to calculate $\mathcal{RF}_{\rm B_{\rm h}}$. We note that the response functions are not evaluated with a constant $\Delta\log\tau_{\rm c}$-grid but rather in a constant $\Delta z$-grid. These plots illustrate that this combination of spectral lines provides a continuous height coverage ranging from the photospheric continuum ($\log\tau_{\rm c}=0$) until the mid chromosphere ($\log\tau_{\rm c} = -6$). Note that the non-LTE lines (all the lines except the Fe {\sc i}) quickly lose temperature sensitivity with height, due to the source function decoupling from the local temperature (see Appendix~\ref{app:betas}). As explained there, this effect is actually more prominent in the Na {\sc i} and Mg {\sc i} lines than in the Ca {\sc ii} lines (see Fig.~\ref{fig:dep_coeffs}). Meanwhile, the sensitivity to the magnetic field and velocity persists almost to $\log\tau_c=-6$. This is because this sensitivity comes predominantly from the line absorption/emission coefficient, which is a local quantity and is not affected by non-LTE effects.

\section{NLTE source functions}
\label{app:source_nlte}

In previous works where the FIRTEZ inversion code was used to analyze spectral lines formed under non-LTE conditions \citep{kaithakkal2023,borrero2024models}, the total source total function $S(\lambda,T)$ for such spectral lines was treated by FIRTEZ using the following expression:

\begin{equation}
S (\lambda) = \frac{2 h c^2}{\lambda^5} \left[\frac{\beta_{\rm l}}{{\beta_{\rm u}}} \exp{\left(\frac{hc}{\lambda KT}\right)}- 1\right]^{-1} \;,
\label{eq:source_before}
\end{equation}

\noindent where $\beta_{\rm u}$ and $\beta_{\rm l}$ correspond to the departure coefficients of the upper and lower levels, respectively, and are defined as the ratio between the LTE and non-LTE level populations. However, note that this expression is only approximate. In reality, Equation~\ref{eq:source_before} is strictly valid only for the line-source-function $S_{\rm l}(\lambda)$ that is only part of the total source function. The other contribution to the total source function is the continuum-source function $S_{\rm c}(\lambda)$ that is always treated under local thermodynamic equilibrium and therefore corresponds to Planck's function $B(\lambda)$. Combining the continuum source function $S_{\rm c}$ and the line source function $S_{\rm l}$ is done by weighting each of them with the ratio of its opacity (continuum or line) to the total (continuum plus line) opacity \citep[see Eq.~14.39 in][]{mihalas2015}:

\begin{equation}
S(\lambda) = \frac{\chi_c}{\chi_c + \sum\limits_{i=0}^{N_{\rm l}} \chi^i_{\rm l}(\lambda)} S_{\rm c}(\lambda) + \sum\limits_{i=0}^{N_{\rm l}} \frac{\chi^i_{\rm l}}{\chi_{\rm c} + \sum\limits_{j=0}^{N_{\rm l}} \chi^j_{\rm l}(\lambda)} S^i_{\rm l} (\lambda)
\label{eq:source_now}
\end{equation}

\noindent where:

\begin{eqnarray}
S_{\rm c} (\lambda) & = & \frac{2 h c^2}{\lambda^5} \left[\exp{\left(\frac{hc}{\lambda KT}\right)}- 1\right]^{-1}\label{eq:source_cont}\\
S_{\rm l} (\lambda) & = & \frac{2 h c^2}{\lambda^5} \left[\frac{\beta_{\rm l}}{{\beta_{\rm u}}} \exp{\left(\frac{hc}{\lambda KT}\right)}- 1\right]^{-1}\label{eq:source_line}
\end{eqnarray}

We note that the line opacity $\chi_{\rm l}$ presents a summation over all possible blending spectral lines: $j=0, ..., N_{\rm l}$. The second term on the right-hand side of Equation~\ref{eq:source_now} also presents a summation over the line source function of all possible blending spectral lines: $S_{\rm l}^{i}$ with $i=0,...,N_{\rm l}$. This new approach included in FIRTEZ allows us to compute, in a more accurate way, the continuum contribution. Another upside is to have different lines blended with independent departure coefficients. Because of this, the current version of FIRTEZ can simultaneously treat any number of blends either on LTE or non-LTE (or a combination of both), whereas in the former version of FIRTEZ, non-LTE spectral lines could not be blended with any other line.\\

Useful for the interpretation of inferred temperature (see Sect.~\ref{subsubsec:temperature}) is to identify the fact that the line-source function becomes close to Planck's function as the ratio between the upper and lower departure coefficients approaches one: $\beta_{\rm u} / \beta_{\rm l} \rightarrow 1$. This is not to say that the level populations are the same as in local-thermodynamic equilibrium, but rather that both upper and lower level populations deviate from LTE in the same way. At visible and near-infrared wavelengths, the exponential in Eq.~\ref{eq:source_line} is typically much larger than 1 and therefore we can write that: $S_{\rm l} \propto \beta_{\rm u} / \beta_{\rm l}$.

\section{Departure coefficients}
\label{app:betas}

To invert the spectra of non-LTE lines, the FIRTEZ code requires departure coefficients of the relevant levels as input. In this case, the Ca{\sc ii}, Na{\sc i}, and Mg{\sc i} lines require non-LTE treatment. Even though the code only requires the departure coefficients of the upper and the lower level of each transition, their calculation requires multi-level atomic models for each. Here we used a 5-level model for Ca {\sc ii} \citep{shine_1974_CaII}, a 5-level model for Na {\sc i} \citep[the first 5 levels from the 18-level model of ][]{Bruls_1992_alkali}, and a 6-level model for Mg {\sc i} created by \citet{vukadinovic_2022_mgIb}. Each of the atomic models has an additional continuum level on the next ionization stage. 
We calculated the level populations for all three species simultaneously using the SNAPI code \citep{Milic_2018_snapi}, taking into account the non-LTE population of electrons calculated using simple charge conservation \citep{Osborne_2021_lw}.\\

\begin{figure}[ht!]
    \centering
    \includegraphics[width=1\linewidth]{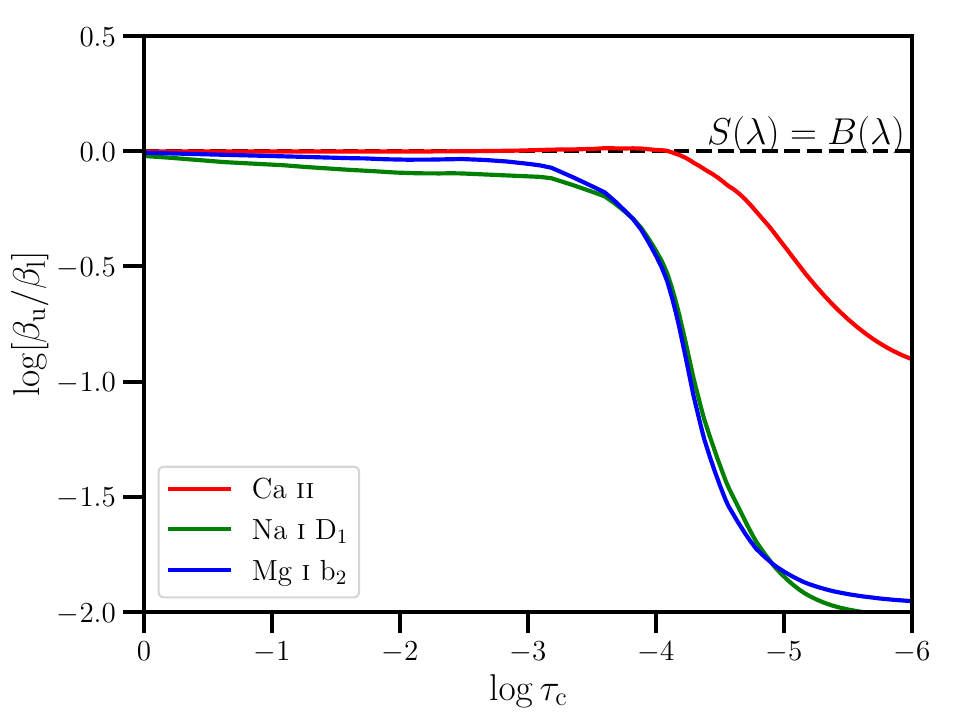}
    \caption{Departure coefficients for non-LTE spectral lines. This plots shows the dependence with optical depth $\log\tau_{\rm c}$ of the ratio of upper and lower level departure coefficients $\beta_{\rm u} / \beta_{\rm l}$. Color lines correspond to each of the analyzed spectral lines formed under non-LTE conditions: Mg {\sc i} b$_2$ (blue), Na {\sc i} D$_1$ (green) and Ca {\sc ii} (red). They were determined for the VALC model and used to initialize the inversion (see Sect.~\ref{subsec:inversion}). The horizontal dashed line indicates the region where $\beta_{\rm u} = \beta_{\rm l}$ and therefore the source function $S(\lambda)$ corresponds to Plank's function $B(\lambda)$ (Sect.~\ref{app:source_nlte}).}
    \label{fig:dep_coeffs}
\end{figure}

Figure~\ref{fig:dep_coeffs} displays the ratio between the upper and lower level departure coefficients (in logarithmic scale) for each of the non-LTE spectral lines analyzed in this work and obtained from the VALC model.
These are the departure coefficients used to initialize the inversion (see Sect.~\ref{subsec:inversion}).
As can be seen, the line source function, being proportional to $\beta_{\rm u} / \beta_{\rm l}$, starts to deviate from LTE in the Na {\sc i} and Mg {\sc i} lines (green and blue lines in Fig.~\ref{fig:dep_coeffs}) already in the upper photosphere ($\tau_{\rm c} < 10^{-3}$) and is completely decoupled from Plank's function, and hence also decoupled from the local temperature, in the chromosphere ($\tau_{\rm c} \approx 10^{-6}$).
Meanwhile, the source function for the Ca {\sc ii} spectral line (red lines in Fig.~\ref{fig:dep_coeffs}) remains close to Planck's function up to the temperature minimum ($\tau_{\rm c} \approx 10^{-4}$) and only starts to deviate from LTE in the chromosphere.\\

We note that the recalculation of the departure coefficients after the first inversion cycle mentioned in Section~\ref{subsec:inversion}, should be done for all observed NLTE lines. However, in practice, this was done only for the Ca {\sc ii} spectral line. The departure coefficients for Mg {\sc i} and Na {\sc i} are kept as those arising from the VALC atmosphere in the entire domain. The reason for this is that the recalculation of new departure for these two lines did not led to better fits of the Stokes parameters, and made the convergence of subsequent inversions harder. Although we do not know with certainty the reason for this, we have identified two possible causes for such behavior.\\

First, the presence of three-dimensional radiative transfer effects. Both Mg {\sc i} b$_2$ and Na {\sc i} D$_1$ are transitions with very high spontaneous emission rate, which means their level populations are dominated by the radiation which can, in principle, come from nearby pixels\citep[e.g.][]{Leenaarts_2010_NaID}. The coupling of neighboring pixels is a consequence of 3D non-LTE radiative transfer, and our calculation of departure coefficients that treats each pixel as a one-dimensional plane-parallel atmosphere cannot account for this coupling. We stress that, while the Ca {\sc ii} line is also a non-LTE sensitive to such effects \citep[e.g.][]{Jaime_2012_inversionsfrom3D}, it is generally more coupled to local conditions than the Na {\sc i} and Mg {\sc i} lines (recall Fig.\,\ref{fig:dep_coeffs}).

A second possible reason for such a discrepancy is that the atomic models used for the Na {\sc i} and Mg {\sc i} lines have been tailored to reproduce the spectral line shapes of the quiet Sun for the VALC atmosphere. Therefore, it is not guaranteed that the re-calculation of the departure coefficients for substantially different atmospheres found in umbra and penumbra will produce the departure coefficients consistent between these two lines, as well as the Ca {\sc ii} line. This effect can be particularly pronounced in the Mg {\sc i} b$_2$ line as the neutral magnesium, as a minority species, requires the so-called opacity fudge \citep[see e.g.,][and references therein]{vukadinovic_2022_mgIb}. The opacity fudge tables are typically calculated for a particular mean atmosphere model, like VALC, and could prove to be inadequate for other atmosphere models. We stress that our work is one of the few studies that attempt to perform a comprehensive multi-line inversion in a sunspot. Sunspots exhibit significantly different physical conditions (temperature, density, etc.) compared to the quiet Sun, so inadequacies of the atomic model or fudge factors are expected to matter more.\\

\end{document}